\documentstyle[preprint,eqsecnum,aps]{revtex}

\def\lesssim{\ \raise.3ex\hbox{$<$}\kern-0.8em\lower.7ex\hbox{$\sim$}\ }
\def\gesim{\ \raise.3ex\hbox{$>$}\kern-0.8em\lower.7ex\hbox{$\sim$}\ }
\font\scripti=cmmi7
\font\scriptscripti=cmmi5
\def\sib#1{\setbox0 = \hbox{\scripti #1}
  \kern-.02em\copy0\kern-\wd0
  \kern.04em\box0} 
\def\ssib#1{\setbox0 = \hbox{\scriptscripti #1}
  \kern-.02em\copy0\kern-\wd0
  \kern.04em\box0} 

\font\tenib=cmmib10 
\skewchar\tenib='177 \skewchar\tenib='177 \skewchar\tenib='177
\def\pbold#1{\setbox0 = \hbox{$ #1 $}
  \kern-.022em\copy0\kern-\wd0
  \kern.011em\copy0\kern-\wd0
  \kern.011em\copy0\kern-\wd0
  \kern.011em\copy0\kern-\wd0
  \kern.011em\box0} 

\begin{document}
\draft
\title{Superfluidity and collective modes in a uniform gas of Fermi atoms with a Feshbach resonance}
\author{Y. Ohashi$^1$ and A. Griffin$^2$}
\address{
$^1$Institute of Physics, University of Tsukuba, Ibaraki 305, Japan \\
$^2$Department of Physics, University of Toronto, Toronto, Ontario, Canada M5S 1A7\\
}
\maketitle
\begin{abstract}
We investigate strong-coupling superfluidity in a uniform two-component gas of ultra-cold Fermi atoms attractively interacting via quasi-molecular bosons associated with a Feshbach resonance. 
This interaction is tunable by the threshold energy $2\nu$ of the Feshbach resonance, becoming large as $2\nu$ is decreased (relative to $2\varepsilon_{\rm F}$, where $\varepsilon_{\rm F}$ is the Fermi energy of one component.). 
In recent work, we showed that the enhancement of this tunable pairing interaction naturally leads to the BCS-BEC crossover, where the character of the superfluid phase transition changes from the BCS-type to a BEC of composite bosons consisting of preformed Cooper-pairs and Feshbach-induced molecules. 
In this paper, we extend our previous work and study both single quasi-particles and the collective dynamics of the superfluid phase below the phase transition temperature $T_{\rm c}$, limiting ourselves to a uniform gas. 
We show how the superfluid order parameter changes from the Cooper-pair amplitude $\Delta$ to the square root of the number of condensed molecules ($\phi_{\rm m}$) associated with the Feshbach resonance, as the threshold energy $2\nu$ is lowered. 
In the intermediate coupling regime, the superfluidity is shown to be characterized by an order parameter consisting of a superposition of $\Delta$ and $\phi_{\rm m}$. We also discuss the Goldstone mode associated with superfluidity, and show how its character smoothly changes from the Anderson-Bogoliubov phonon in the BCS regime to the Bogoliubov phonon in the BEC regime in the BCS-BEC crossover. The velocity of this Goldstone phonon mode is shown to strongly depend on the value of $2\nu$. We also show that this Goldstone mode appears as a resonance in the spectrum of the density-density correlation function, which is experimentally accessible. 
\par
\vskip10mm
\end{abstract}
\vskip3mm
\pacs{PACS numbers: 03.75.Kk, 03.75.Ss, 74.20.Mn}
\narrowtext
%
\section{Introduction}
One of the most challenging topics in current physics is the realization of superfluidity in a trapped atomic gas composed of two Fermi hyperfine states. A considerable experimental effort has already been made to cool down Fermi atom gases, such as $^6$Li and $^{40}$K\cite{Jin,Andrew,Salomon,Granade,Inguscio}. The temperature can now be lowered to $T\sim 0.2T_{\rm F}$, where the Fermi gas should be highly degenerate and the observation of superfluid behavior seems imminent\cite{Thomas}.
\par
As a promising mechanism of superfluidity with a high phase transition temperature $T_{\rm c}$, making use of an atomic Feshbach resonance has attracted much attention\cite{Timmermans1,Timmermans2,Holland,Chiofalo,Kokkelman,Ohashi1,Ohashi2,Ohashi3,Milstein}. The Feshbach resonance describes quasi-molecular bosons, which can mediate a pairing interaction between Fermi atoms. This pairing interaction is tunable by the threshold energy $2\nu$ of the Feshbach resonance, and can become strong as $2\nu$ is decreased relative to twice the Fermi energy of the atoms. 
Using this strong paring interaction, one can hope to achieve a high value of $T_{\rm c}$. Experimentally, the threshold energy $2\nu$ can be controlled by a weak applied magnetic field. Very recently, this tunable interaction was observed in a Fermi gas of $^{40}$K\cite{Loftus,Regel}. 
\par
In our recent work\cite{Ohashi1,Ohashi2,Ohashi3}, we pointed out the importance of fluctuations in the Cooper-channel in considering a high-$T_{\rm c}$ superfluidity originating from the strong pairing interaction associated with the Feshbach resonance. We extended the strong-coupling theory developed by Nozi\`eres and Schmitt-Rink\cite{Nozieres,Tokumitsu,Melo,Randeria1}, to include the effects of a Feshbach resonance and the associated quasi-molecular bosons. We showed that these particle-particle fluctuations strongly suppress $T_{\rm c}$ from the value expected within the simple mean-field BCS theory. In addition, the character of the phase transition was shown to continuously change from the BCS-type to a BEC of composite bosons (consisting of preformed Cooper-pairs and long-lived Feshbach molecules) as the threshold energy $2\nu$ is lowered. Our strong-coupling theory thus gave an upper limit of $T_{\rm c}=0.518T_{\rm F}$ for a Fermi gas in an isotropic harmonic trap potential and $T_{\rm c}=0.218T_{\rm F}$ for a uniform gas\cite{Ohashi1,Ohashi2,Ohashi3}.
These values are simply the BEC transition temperatures, expressed in terms of the Fermi temperature $T_{\rm F}$ of one of the Fermi components.
\par
In this paper, we investigate the BCS-BEC crossover in the superfluid state, extending our previous work\cite{Ohashi1,Ohashi2,Ohashi3} to the superfluid region below $T_{\rm c}$. Going past the previous BCS mean-field approximation\cite{Holland,Lee,Ranninger1,Ranninger2}, we include strong fluctuations around the BCS mean-field solution. We clarify how the order parameter described by the Cooper-pair amplitude $\Delta=U\sum_{\sib p}\langle c_{-{\sib p}\downarrow}c_{{\sib p}\uparrow}\rangle$ in the weak-coupling BCS theory changes to the BEC order parameter related to the number of condensed bosons $\phi_{\rm m}=\langle b_{{\sib q}=0}\rangle$ in the BCS-BEC crossover. Here $c_{{\sib p}\sigma}$ is the annihilation operator of a Fermi atom in one of two hyperfine states ($\sigma=\uparrow,\downarrow$) and $b_{\sib q}$ is the annihilation operator of the boson molecule associated with the Feshbach resonance.
\par
In the field of trapped ultra-cold Fermi gases, a crucial issue is to determine a clear unambiguous signature for superfluidity\cite{Thomas,Weig,Bruun1,Bruun2,Stingare}. Another important problem is how to experimentally track the system in the BCS-BEC crossover region. In this regard, the Goldstone collective mode is a very useful quantity because it is deeply related to the spontaneous breakdown of the gauge symmetry associated with the superfluid phase transition. The Goldstone mode is known as the Anderson-Bogoliubov mode in the BCS state\cite{Anderson}, while it is the Bogoliubov phonon in the BEC phase\cite{Bogoliubov}. In this paper, we discuss how these collective modes change from one to the other as we go through the BCS-BEC crossover. 
We show that the velocity of the Goldstone phonon $v_\phi$ strongly depends on $2\nu$, and thus it offers a way of observing the BCS-BEC crossover phenomenon by tuning the threshold energy $2\nu$ in a cigar shaped trap (where the gas is fairly uniform in the axial direction).
We also show that the Goldstone mode appears as a resonance in the spectrum of the density-density correlation function.
\par
The present paper only considers the superfluid phase in a uniform two-component Fermi gas with an attractive interaction. In Ref. \cite{Ohashi3}, we discussed the same model at and above the superfluid transition temperature for a trapped gas, using the local density approximation (LDA). The extension of the present work to an inhomogeneous trapped gas will be considered in the future. As well known, the low energy collective modes in a trapped atomic gas (Fermi or Bose) are strongly altered by the trap potential. However, the equivalent of the Goldstone phonon modes we discuss in this paper also arise in trapped two component Fermi gases. These low energy collective modes have been extensively discussed in the recent literature\cite{Baranov,Mot,Bruun3} for the weak-coupling BCS limit. In this paper, we discuss the physics of the
Goldstone phonon mode as a function of the threshold energy $2\nu$. A similar analysis remains to be done for the collective modes of a trapped Fermi gas in the BCS-BEC crossover region.
\par
This paper is organized as follows. In Section II, we present our coupled fermion-boson model. We explain how to include the strong-coupling (fluctuation) effect originating from the Feshbach resonance in Section III. In Section IV, we consider the strong-coupling theory presented in Section III in terms of Gaussian fluctuations. 
In Section V, we consider the Goldstone mode. We first derive correlation functions describing Cooper-pair fluctuations, as well as a renormalized boson Green's function for quasi-molecules associated with the Feshbach resonance, under the Hartree-Fock random phase approximation (HF-RPA). The Goldstone mode is then obtained from their poles. 
In Section VI, we discuss the BCS-BEC crossover below $T_{\rm c}$ based on our numerical results. In this section, we also discuss the BCS-BEC crossover behavior of the order parameter and the Goldstone phonon mode, as a function of $2\nu$. In Section VII, we discuss the coupling of the Goldstone mode with the density fluctuations in a gas of Fermi atoms.
\par
\vskip3mm
\section{Coupled fermion-boson model}
We consider a gas of Fermi atoms composed of two atomic hyperfine states, coupled to molecular two particle state. We describe the two hyperfine states using a pseudo-spin variable $\sigma$ ($=\uparrow,~\downarrow$). The coupled fermion-boson model Hamiltonian is given by\cite{Timmermans1,Timmermans2,Holland,Chiofalo,Kokkelman,Ohashi1,Ohashi2,Ohashi3,Milstein,Lee,Ranninger1,Ranninger2}
\begin{eqnarray}
{\cal H}
&=&
\sum_{{\sib p}\sigma}\varepsilon_{\sib p}
c_{{\sib p}\sigma}^\dagger c_{{\sib p}\sigma}
-
U\sum_{{\sib p},{\sib p}',{\sib q}}
c^\dagger_{{\sib p}+{{\sib q} \over 2}\uparrow}
c^\dagger_{-{\sib p}+{{\sib q} \over 2}\downarrow}
c_{-{\sib p}'+{{\sib q} \over 2}\downarrow}
c_{{\sib p}'+{{\sib q} \over 2}\uparrow}
\nonumber
\\
&+&
\sum_{\sib q}[\varepsilon_{B\sib q}+2\nu] b_{\sib q}^\dagger b_{\sib q}
+
g_{\rm r}\sum_{{\sib p},{\sib q}}
[
b_{\sib q}^\dagger 
c_{-{\sib p}+{{\sib q} \over 2}\downarrow}
c_{{\sib p}+{{\sib q} \over 2}\uparrow}
+
{\rm h.c.}
].
\label{eq.2.1}
\end{eqnarray}
Here a Fermi atom and a quasi-molecular boson associated with the Feshbach resonance are, respectively, described by the destruction operators $c_{{\sib p}\sigma}$ and $b_{\sib q}$. The kinetic energy of a Fermi atom is $\varepsilon_{\sib p}\equiv p^2/2m$, and $\varepsilon_{B\sib q}+2\nu\equiv q^2/2M+2\nu$ is the excitation spectrum of the $b$-molecular bosons. Here $2\nu$ represents the lowest excitation energy of $b$-bosons, also referred to as the threshold energy of the Feshbach resonance. The last term in Eq. (\ref{eq.2.1}) describes the Feshbach resonance with a coupling constant $g_{\rm r}$, which describes how a $b$-molecule can dissociate into two Fermi atoms and how two Fermi atoms can form one $b$-boson. The Hamiltonian in Eq. (\ref{eq.2.1}) also includes an attractive fermion-fermion interaction $-U$ ($<0$) arising from non-resonant processes\cite{Holland}. 
\par
Since one $b$-Bose molecule consists of two Fermi atoms, the boson mass $M=2m$ and the conservation of the total number of particles $N$ imposes the relation 
\begin{equation}
N=\sum_{{\sib p}\sigma}
\langle c^\dagger_{{\sib p}\sigma}c_{{\sib p}\sigma}\rangle
+
2\sum_{\sib q}
\langle b_{\sib q}^\dagger b_{\sib q}\rangle.
\label{eq.2.2}
\end{equation}
We incorporate this latter constraint into the model Hamiltonian in Eq. (\ref{eq.2.1}) using a chemical potential, ${\cal H}\equiv H-\mu N$. The resulting grand-canonical Hamiltonian ${\cal H}$ has the same form as Eq. (\ref{eq.2.1}), except that the kinetic energies of Fermi atoms and $b$-bosons are replaced as $\varepsilon_{\sib p}\to\xi_{\sib p}\equiv\varepsilon_{\sib p}-\mu$ and $\varepsilon_{B\sib q}+2\nu\to\xi_{B\sib q}\equiv\varepsilon_{B\sib q}+2\nu-2\mu$, respectively. In the latter replacement, the factor $2$ in $2\mu$ reflects that one $b$-boson consists of two Fermi atoms. 
\par
In this paper, we investigate strong-coupling effects in the superfluid phase, as well as the Goldstone mode associated with superfluidity in the BCS-BEC crossover region. As a start, we consider a simple uniform Fermi gas and leave the effect of a trapping potential to future work. In this regard, we have shown in Refs.\cite{Ohashi1,Ohashi2,Ohashi3}. that while a trap potential enhances the transition temperature $T_{\rm c}$ in the BEC regime, the qualitative behavior of $T_{\rm c}$ in the BCS-BEC crossover is not very different from a uniform Fermi gas. Within weak-coupling BCS theory, several papers have discussed collective excitations in a trapped Fermi gas with attractive interactions. (See, for example, Refs. \cite{Baranov,Mot,Bruun3}.)
\par
When the Feshbach coupling term is absent in Eq. (\ref{eq.2.1}), the fermions and $b$-bosons are decoupled from each other. In this limit, a BCS superfluid phase transition of Fermi atoms and BEC transition of $b$-bosons can occur, at different temperatures. These two superfluid phases are, respectively, described by independent order parameters
\begin{eqnarray}
\left\{
\begin{array}{l}
\Delta\equiv
U\sum_{\sib p}\langle c_{-{\sib p}\downarrow}c_{{\sib p}\uparrow}\rangle,
\\
\phi_{\rm m}\equiv\langle b_{{\sib q}=0} \rangle.
\end{array}
\right.
\label{eq.2.3}
\end{eqnarray}
On the other hand, when the Feshbach resonance term is present ($g_{\rm r}\ne 0$), we find the following identity in the equilibrium state:
\begin{eqnarray}
0
=i{d \phi_{\rm m}\over dt}
=i\Bigl\langle{d b_0 \over dt}\Bigr\rangle
=\langle[b_0,{\cal H}]\rangle
=(2\nu-2\mu)\phi_{\rm m}+{g_{\rm r} \over U}\Delta.
\label{eq.2.4}
\end{eqnarray}
Eq. (\ref{eq.2.4}) gives\cite{Timmermans2,Holland}
\begin{equation}
\phi_{\rm m}=-{g_{\rm r} \over 2\nu-2\mu}{\Delta \over U}.
\label{eq.2.5}
\end{equation}
This last result shows that the BEC order parameter $\phi_{\rm m}$ and the Cooper-pair order parameter $\Delta$ are no longer independent, due to the hybridization induced by the Feshbach resonance $g_{\rm r}$. Both $\Delta$ and $\phi_{\rm m}$ are finite in the superfluid phase, and there is a unique superfluid phase transition in this coupled fermion-boson model.
\par
For later convenience, we define the following composite order parameter\cite{Timmermans2,Holland,Ohashi1}
\begin{equation}
{\tilde \Delta}\equiv\Delta-g_{\rm r}\phi_{\rm m}.
\label{eq.2.6}
\end{equation}
We will find that ${\tilde \Delta}$ determines the excitation energy gap in the spectrum of fermion quasi-particles below $T_{\rm c}$ in the BCS-BEC crossover regime.
\vskip3mm
\section{Strong-coupling effect on superfluidity}
\vskip2mm
\subsubsection{Review on the strong-coupling theory for $T_{\rm c}$}
\vskip2mm
In this section, we review the strong-coupling theory for $T_{\rm c}$ discussed in our previous papers\cite{Ohashi1,Ohashi2,Ohashi3}. This formulation is extended to the region below $T_{\rm c}$ in the next subsection. 
\par
In previous work\cite{Ohashi1,Ohashi2,Ohashi3}, we extended the strong-coupling theory developed by Nozi\`eres and Schmitt-Rink\cite{Nozieres} to the coupled fermion-boson model in Eq. (\ref{eq.2.1}). The equation for $T_{\rm c}$ was obtained by using the Thouless criterion, which states that the superfluid phase transition occurs when the particle-particle vertex function $\Gamma({\bf q},\omega)$ describing the Cooper-channel develops a pole at ${\bf q}={\omega}=0$. Within the $t$-matrix approximation in terms of $-U$ and $g_{\rm r}$ described diagrammatically in Fig. 1(a), this equation for $T_{\rm c}$ is given by
\begin{equation}
1=U_{\rm eff}\sum_{\sib p}{\tanh 2\xi_{\sib p}/2T_{\rm c} \over 2\xi_{\sib p}},
\label{eq.2.7}
\end{equation}
where 
\begin{equation}
U_{\rm eff}\equiv U+g_{\rm r}^2 {1 \over 2\nu-2\mu}
\label{eq.2.8}
\end{equation}
is an effective pairing interaction. The last term in Eq. (\ref{eq.2.8}) describes the interaction mediated by $b$-bosons associated with the Feshbach resonance. Eq. (\ref{eq.2.7}) is formally identical to the equation for $T_{\rm c}$ in ordinary weak-coupling BCS theory. However, the chemical potential $\mu$ in the kinetic energy $\xi_{\sib p}=\varepsilon_{\sib p}-\mu$ of the Fermi atoms can deviate strongly from $\varepsilon_{\rm F}$ as one approaches the BEC regime (where $\varepsilon_{\rm F}$ is the bare Fermi energy of one spin component). This contrasts with simple BCS theory, where one finds that $\mu\simeq\varepsilon_{\rm F}$. 
\par
The chemical potential $\mu$ is determined by the equation for the total number of Fermi atoms, using the identity $N=-\partial\Omega/\partial\mu$. We include the effect of fluctuations in the Cooper-channel [the first line in Fig. 1(b)] as well as the Feshbach resonance [the second line in Fig. 1(b)] in the thermodynamic potential $\Omega$. The resulting equation relating $\mu$ and $N$ is
\begin{equation}
N=N_{\rm F}^0+2N_{\rm B}^0-{1 \over \beta}\sum_{{\sib q},\nu_n}
e^{i\delta\nu_n}
{\partial \over \partial\mu}
\ln
\Bigl[
1-[U-g_{\rm r}^2D_0({\bf q},i\nu_n)]\Pi({\bf q},i\nu_n)
\Bigr].
\label{eq.2.9}
\end{equation} 
Here $\beta=1/T$ is the inverse of temperature, $N^0_{\rm F}\equiv2\sum_{\sib p}f(\xi_{\sib p})$ and $N^0_{\rm B}\equiv\sum_{\sib q} n_{\rm B}(\xi_{B\sib q})$, where $f(\varepsilon)$ and $n_{\rm B}(\varepsilon)$ represent the Fermi and Bose distribution functions, respectively. The last term in Eq. (\ref{eq.2.9}) describes the fluctuation contribution to $N$. The $b$-boson Green's function is given by 
\begin{equation}
D_0({\bf r},i\nu_n)\equiv{1 \over i\nu_n-\xi_{B\sib q}},
\label{eq.2.10}
\end{equation}
where the Bose Matsubara frequency is $\nu_n=2n\pi T~(n=0,\pm1,\pm2,\cdot\cdot\cdot)$. $\Pi({\bf q},i\nu_n)$ is the correlation function of the Cooper-pair field operator ${\hat F}_{\sib q}\equiv\sum_{\sib p}c_{-{\sib p}+{{\sib q} \over 2}\downarrow} c_{{\sib p}+{{\sib q} \over 2}\uparrow}$, given by
\begin{eqnarray}
\Pi({\bf q},i\nu_n)
&\equiv&
\int_0^\beta d\tau e^{i\nu_n\tau}
\langle
T_\tau
\{
{\hat F}_{\sib q}(\tau){\hat F}_{\sib q}^\dagger(0)
\}
\rangle
\nonumber
\\
&=&
{1 \over \beta}
\sum_{{\sib p},\omega_m}
G_0({\bf p}+{{\bf q} \over 2},i\omega_m+i\nu_n)
G_0(-{\bf p}+{{\bf q} \over 2},i\omega_m)
\nonumber
\\
&=&
\sum_{\sib p}
{
1-f(\xi_{{\sib p}+{{\sib q} \over 2}})-f(\xi_{{\sib p}-{{\sib q} \over 2}})
\over
\xi_{{\sib p}+{{\sib q} \over 2}}+\xi_{{\sib p}-{{\sib q} \over 2}}-i\nu_n
}.
\label{eq.2.11}
\end{eqnarray}
Physically, $\Pi({\bf q},i\nu_n)$ describes fluctuations of Cooper-pairs in the normal phase. $G_0({\bf q},i\omega_m)$ is a fermion thermal Green's function defined by
\begin{equation}
G_0({\bf q},i\omega_m)={1 \over i\omega_m-\xi_{\sib p}},
\label{eq.2.12}
\end{equation}
where the Fermi Matsubara frequency is $\omega_m=(2m+1)\pi T~(m=0,\pm1,\pm2,\cdot\cdot\cdot)$. 
\par
The coupled equations (\ref{eq.2.7}) and (\ref{eq.2.9}) determine $T_{\rm c}$ and $\mu$ self-consistently. In calculating these equations, a cutoff is necessary to make the momentum summation converge. In Refs.\cite{Ohashi1,Ohashi2,Ohashi3}, we simply introduced a Gaussian cutoff $e^{-\varepsilon_{\sib p}^2/\omega_c^2}$, which is also used in this paper.
\par
The self-consistent solution ($T_{\rm c}$ and $\mu$) of these coupled equations is summarized in Fig.2. When $\nu\gg\varepsilon_{\rm F}$, since the chemical potential is at most $\mu\lesssim \varepsilon_{\rm F}$, the Feshbach-induced contribution to the pairing interaction $g_{\rm r}^2/(2\nu-2\mu)$ in Eq. (\ref{eq.2.8}) is small. In this regime, Fig. 2(a) shows that the superfluid phase transition is well described by the weak-coupling BCS theory with a (weak) pairing interaction $-U$. In addition, we see that $\mu\simeq\varepsilon_{\rm F}$, as shown in Fig. 2(b). However, the chemical potential $\mu$ gradually deviates from $\varepsilon_{\rm F}$ as the threshold energy $2\nu$ is lowered towards $2\varepsilon_{\rm F}$ and below. 
In particular, one finds $\mu$ approaches $\nu$ when $\nu<0$. 
In this regime, the Feshbach-induced pairing interaction $g_{\rm r}^2/(2\nu-2\mu)$ in Eq. (\ref{eq.2.8}) is large, and $T_{\rm c}$ deviates significantly from the prediction of weak-coupling BCS theory, as shown in Fig. 2(a). Since $2\nu$ is the lowest excitation energy of $b$-bosons and their chemical potential is $2\mu$, the situation $2\nu=2\mu$, realized in the limit of large negative values of $\nu/\varepsilon_{\rm F}$, is equivalent to the condition of BEC in a non-interacting Bose gas. Indeed, Fig. 2(a) shows that $T_{\rm c}$ corresponds precisely to the transition temperature of a free Bose gas of $N/2$ atoms when $\nu/\varepsilon_{\rm F}<-1$. These results show that the BCS-BEC crossover occurs in the region around zero threshold ($2\nu=0$), at least for small values of $g_{\rm r}$. This crossover phenomenon can be simply controlled by the threshold energy $2\nu$ of the Feshbach resonance.
\vskip3mm
\subsubsection{Strong-coupling theory below $T_{\rm c}$}
\vskip2mm
In order to formulate the analogous strong-coupling theory below $T_{\rm c}$, we separate out the fluctuations of Cooper-pairs and condensed $b$-bosons around their mean-field values denoted by $\Delta$ and $\phi_{\rm m}$, respectively\cite{OhashiT1,OhashiT2}. For this purpose, we write ${\hat F}_{{\sib q}}=\langle {\hat F}_{{\sib q}=0}\rangle\delta_{{\sib q},0}+\delta {\hat F}_{{\sib q}}$ and ${\hat b}_{{\sib q}}=\phi_m\delta_{{\sib q},0}+\delta b_{{\sib q}}$, where ${\hat F}_{{\sib q}}$ is defined before Eq. (\ref{eq.2.11}). Separating out the fluctuation contribution to the Hamiltonian in Eq. (\ref{eq.2.1}), we obtain 
\begin{eqnarray}
{\cal H}
&=&
{{\tilde \Delta}^2 \over U_{\rm eff}}+\sum_{\sib p}\xi_{\sib p}
+\sum_{\sib p}
{\hat \Psi}^\dagger_{\sib p}
[\xi_{\sib p}\tau_3-{\tilde \Delta}\tau_1]
{\hat \Psi}_{\sib p}+
\sum_{\sib q}\xi_{B\sib q}b_{\sib q}^\dagger b_{\sib q}
\nonumber
\\
&+&
{g_{\rm r} \over 2}\sum_{\sib q}
[b^\dagger_{\sib q}\rho_{\sib q}^-+b_{\sib q}\rho_{\sib q}^+]
-{U \over 4}
\sum_{\sib q}[\rho_{1{\sib q}}\rho_{1-{\sib q}}+\rho_{2{\sib q}}\rho_{2-{\sib q}}].
\label{eq.2.13}
\end{eqnarray}
Here we have introduced the Nambu field operator for Fermi atoms as ${\hat \Psi}^\dagger_{\sib p}\equiv (c_{{\sib p}\uparrow}^\dagger,c_{-{\sib p}\downarrow})$ and the corresponding $2\times 2$-Pauli matrices $\tau_i$ ($i=1,2,3$) acting on the particle-hole space\cite{Sch}. The order parameter ${\tilde \Delta}$ is defined by Eqs. (\ref{eq.2.3}) and (\ref{eq.2.6}), which we can take to be real and proportional to 
$\tau_1=
\left(
\begin{array}{cc}
0 & 1 \\ 1& 0
\end{array}
\right)
$
 without loss of generality. In Eq. (\ref{eq.2.13}) and the subsequent discussion, we use the generalized density operators $\rho_{\sib q}^\pm\equiv\rho_{1\pm{\sib q}}\pm i\rho_{2\pm{\sib q}}$, where
\begin{equation}
\rho_{j\sib q}\equiv
\sum_{\sib p}\Psi^\dagger_{{\sib p}+{{\sib q} \over 2}}
\tau_j\Psi_{{\sib p}-{{\sib q} \over 2}}.
\label{eq.2.14}
\end{equation}
We note that $\rho_{3{\sib q}}=\sum_{{\sib p},\sigma}c^\dagger_{{\sib p}+{{\sib q} \over 2},\sigma}c_{{\sib p}-{{\sib q} \over 2},\sigma}$ is the ordinary density fluctuation operator. Similarly, one has
\begin{eqnarray}
\left\{
\begin{array}{l}
\rho_{1{\sib q}}=\sum_{\sib p}
[
c_{-{\sib p}-{{\sib q} \over 2}\downarrow}
c_{{\sib p}-{{\sib q} \over 2}\uparrow}
+
c^\dagger_{{\sib p}+{{\sib q} \over 2}\uparrow}
c^\dagger_{-{\sib p}+{{\sib q} \over 2}\downarrow}
],
\\
\rho_{2{\sib q}}=i\sum_{\sib p}
[
c_{-{\sib p}-{{\sib q} \over 2}\downarrow}
c_{{\sib p}-{{\sib q} \over 2}\uparrow}
-
c^\dagger_{{\sib p}+{{\sib q} \over 2}\uparrow}
c^\dagger_{-{\sib p}+{{\sib q} \over 2}\downarrow}
],
\end{array}
\right.
\label{eq.AG1}
\end{eqnarray}
and hence
\begin{eqnarray}
\left\{
\begin{array}{l}
\rho^+_{{\sib q}}=2\sum_{\sib p}
c^\dagger_{{\sib p}+{{\sib q} \over 2}\uparrow}
c^\dagger_{-{\sib p}+{{\sib q} \over 2}\downarrow}
,
\\
\rho^-_{{\sib q}}=2\sum_{\sib p}
c_{-{\sib p}-{{\sib q} \over 2}\downarrow}
c_{{\sib p}-{{\sib q} \over 2}\uparrow}.
\end{array}
\right.
\label{eq.AG.2}
\end{eqnarray}
The operators $\rho_{1{\sib q}}$ and $\rho_{2{\sib q}}$ describe, respectively, the amplitude and phase fluctuations of Cooper-pair field fluctuation operator ${\hat F}_{\sib q}$. In Eq. (\ref{eq.2.13}), the fermion-fermion interaction is seen to be neatly expressed as the sum of interactions between the amplitude fluctuations ($-{U \over 4}\sum_{\sib q}\rho_{1{\sib q}}\rho_{1-{\sib q}}$) and phase fluctuations ($-{U \over 4}\sum_{\sib q}\rho_{2{\sib q}}\rho_{2-{\sib q}}$). The Feshbach resonance is also expressed as an interaction between the $b$-bosons and the fluctuations described by $\rho^\pm_{\sib q}$. In Eq. (\ref{eq.2.13}), we have simply written $\rho_{1{\sib q}=0}-\langle\rho_{1{\sib q}=0}\rangle\to\rho_{1{\sib q}=0}$ and $b_{{\sib q}=0}-\phi_{\rm m}\to b_{{\sib q}=0}$. 
\par
Within the mean-field approximation described by the third term in Eq. (\ref{eq.2.13}), the fermion thermal Green's function is conveniently discussed in terms of a $2\times 2$ matrix Green's function
\begin{equation}
{\hat G}({\bf p},i\omega_m)={1 \over i\omega_m-\xi_{\sib p}\tau_3+{\tilde \Delta}\tau_1}
=
-
{
i\omega_m+\xi_{\sib p}\tau_3-{\tilde \Delta}\tau_1
\over 
\omega_m^2+E_{\sib p}^2
}
.
\label{eq.2.15}
\end{equation}
Here $E_{\sib p}\equiv\sqrt{\xi_{\sib p}^2+{\tilde \Delta}^2}$ is the energy spectrum of fermion quasi-particles below $T_{\rm c}$, which we shall call the BCS-Bogoliubov quasi-particle spectrum. 
Eq. (\ref{eq.2.15}) reduces to the mean-field BCS matrix Green's function\cite{Sch} when the composite order parameter ${\tilde \Delta}$ is replaced by $\Delta$. The off-diagonal (static) self-energy ${\hat \Sigma}_{\rm F}\equiv-{\tilde \Delta}\tau_1$ comes from the mean-field term ${\tilde \Delta}\tau_1$ appearing in the Hamiltonian in Eq. (\ref{eq.2.13}). This self-energy corresponds to the mean-field diagrams shown in Fig.3, 
\begin{eqnarray}
\left\{
\begin{array}{l}
\displaystyle
{\hat \Sigma}_{\rm (a)}=
-\tau_1{U \over 2\beta}
\sum_{{\sib p},\omega_m}
{\rm Tr}[\tau_1{\hat G}({\bf p},i\omega_m)]
=
-\tau_1U
\sum_{\sib p}
{{\tilde \Delta} \over 2E_{\sib p}}\tanh{\beta \over 2}E_{\sib p},
\\
\displaystyle
{\hat \Sigma}_{\rm (b)}=
-\tau_+{g_{\rm r}^2 \over 4\beta}D_0(0,0)
\sum_{{\sib p},\omega_m}
{\rm Tr}[\tau_-{\hat G}({\bf p},i\omega_m)]
=
-\tau_+
{g_{\rm r}^2 \over 2\nu-2\mu}
\sum_{\sib p}
{{\tilde \Delta} \over 4E_{\sib p}}\tanh{\beta \over 2}E_{\sib p},
\\
\displaystyle
{\hat \Sigma}_{\rm (c)}=
-\tau_-{g_{\rm r}^2 \over 4\beta}D_0(0,0)
\sum_{{\sib p},\omega_m}
{\rm Tr}[\tau_+{\hat G}({\bf p},i\omega_m)]
=
-\tau_-
{g_{\rm r}^2 \over 2\nu-2\mu}
\sum_{\sib p}
{{\tilde \Delta} \over 4E_{\sib p}}\tanh{\beta \over 2}E_{\sib p},
\end{array}
\right.
\label{eq.AG.100}
\end{eqnarray}
where $\tau_\pm\equiv\tau_1\pm i\tau_2$. In Eq. (\ref{eq.AG.100}), ${\hat \Sigma}_{\rm (b)}$ and ${\hat \Sigma}_{\rm (c)}$ include the pairing interaction mediated by a Feshbach $b$-molecule described by the propagator $D_0$. ${\hat \Sigma}_{\rm (a)}$ comes from the (weak) non-resonant interaction $-U$. The matrix self-energy ${\hat \Sigma}_{\rm F}={\hat \Sigma}_{\rm (a)}+{\hat \Sigma}_{\rm (b)}+{\hat \Sigma}_{\rm (c)}$ sums up to give
\begin{equation}
{\hat \Sigma}_{\rm F}=-\tau_1U_{\rm eff}\sum_{\sib p}
{{\tilde \Delta} \over 2E_{\sib p}}
\tanh{\beta \over 2}E_{\sib p}=-\tau_1{\tilde \Delta}.
\label{eq.AG.101}
\end{equation}
In the last expression, we have used the gap equation in Eq. (\ref{eq.2.17}).
\par
Using eq. (\ref{eq.2.15}), we see that
\begin{eqnarray}
\Delta\equiv U\sum_{\sib p}\langle 
c_{-{\sib p}\downarrow}c_{{\sib p}\uparrow}\rangle=
{U \over 2\beta}\sum_{{\sib p},\omega_m}{\rm Tr}[\tau_1 {\hat G}({\bf p},i\omega_m)]
=U\sum_{\sib p}{{\tilde \Delta} \over 2E_{\sib p}}\tanh{\beta \over 2} E_{\sib p}.
\label{eq.2.16}
\end{eqnarray}

The order parameters $\phi_{\rm m}$ and $\Delta$ can be obtained in terms of ${\tilde \Delta}$ by using Eqs. (\ref{eq.2.5}) and (\ref{eq.2.6}), to give
\begin{eqnarray}
\left\{
\begin{array}{l}
\displaystyle
\Delta={U \over U_{\rm eff}}{\tilde \Delta},
\\
\displaystyle
\phi_{\rm m}=-{g_{\rm r} \over 2\nu-2\mu}{1 \over U_{\rm eff}}{\tilde \Delta}.
\end{array}
\right.
\label{eq.2.18}
\end{eqnarray}
The equation for the composite order parameter ${\tilde \Delta}=\Delta-g_{\rm r}\phi_{\rm m}$ can then be rewritten in the form
\begin{equation}
{\tilde \Delta}=U_{\rm eff}\sum_{\sib p}{{\tilde \Delta} \over 2E_{\sib p}}\tanh{\beta \over 2}E_{\sib p},
\label{eq.2.17}
\end{equation}
where $U_{\rm eff}$ is defined in Eq. (\ref{eq.2.8}). This self-consistent equation has the same form as the BCS gap equation if we replace ${\tilde \Delta}\to\Delta$ and $\mu\to\varepsilon_{\rm F}$. We also note that Eq. (\ref{eq.2.17}) reduces to the $T_{\rm c}$ equation in Eq. (\ref{eq.2.7}) when ${\tilde \Delta}\to 0$\cite{Ohashi1}. 
\par
It is important to emphasize that the $2\times 2$-matrix single-particle Green's function in Eq. (\ref{eq.2.15}) only includes self-energy effects arising from the (off-diagonal) static mean-fields produced by the Cooper-pairs and the Bose-condensed $b$-molecules. In the approximation we use in this paper, the frequency-dependent fermion self-energies associated with the order parameter collective modes are not included in Eq. (\ref{eq.2.15}). However, Eq. (\ref{eq.2.15}) does implicitly involve the self-consistent renormalized values of ${\tilde \Delta}$ and $\mu$, as determined by the order parameter fluctuations (for further discussion of this kind of approximation, see Ref.\cite{Cote}). An improved theory of the BCS-BEC crossover would be based on including the fermion self-energies arising from coupling to collective modes\cite{Haussmann2}. 
\par
The chemical potential $\mu$ is determined from the equation for the total number of particles $N$. As in our discussion of the strong-coupling theory for $T_{\rm c}$\cite{Ohashi1,Ohashi2,Ohashi3}, we work with the thermodynamic potential $\Omega$ consisting of a static mean-field part ($\equiv\Omega_{\rm MF}$) and a fluctuation part ($\equiv\delta\Omega$) originating from the particle-particle Cooper-channel, as modified by the Feshbach resonance. The self-consistent equation for $N$ is given using the identity $N=-\partial\Omega/\partial\mu$. The mean-field part is easily obtained from the first four terms on the right hand side in Eq. (\ref{eq.2.13})\cite{note1},
\begin{equation}
\Omega_{\rm MF}=
{{\tilde \Delta}^2 \over U_{\rm eff}}
+\sum_{\sib p}
(\xi_{\sib p}-E_{\sib p})
-2T
\sum_{\sib p}\ln\Bigl[1+e^{-\beta E_{\sib p}}\Bigr]
+T\sum_{\sib q}\ln\Bigl[1-e^{-\beta\xi_{B\sib q}}\Bigr].
\label{eq.2.19}
\end{equation}
\par
Our approximation for the fluctuation contribution $\delta\Omega$ below $T_{\rm c}$, corresponding to the contribution at $T_{\rm c}$ shown in Fig. 1(b), is given diagrammatically in Fig. 4(a). Among the interaction terms in Eq. (\ref{eq.2.13}), we first carry out perturbative expansion in terms of $-{U \over 4}\sum_{\sib q}\rho_{2{\sib q}}\rho_{2-{\sib q}}$, which describes the interaction between the phase fluctuations of $\Delta$. Summing up the loop-type diagrams in Fig. 4(a), we obtain the phase fluctuation contribution to $\delta\Omega$
\begin{equation}
\delta\Omega_2={1 \over 2\beta}\sum_{{\sib q},\nu_n}
\ln
\Bigl[
1+{U \over 2}\Pi^0_{22}({\bf q},i\nu_n)
\Bigr].
\label{eq.2.20}
\end{equation}
In this $\nu_n$-summation and in the following equations, we omit the important convergence factor $e^{i\nu_n\delta}$, for simplicity of notation. The generalized density correlation function $\Pi^0_{22}({\bf q},i\nu_n)$ is defined by\cite{OhashiT1,OhashiT2}
\begin{eqnarray}
\Pi_{ij}({\bf q},i\nu_n)
&=&
-\int_0^\beta d\tau e^{i\nu_n\tau}
\langle T_\tau\{
\rho_{i{\sib q}}(\tau)\rho_{j-{\sib q}}(0)
\}\rangle~~~(i,j=1,2,3)
\nonumber
\\
&=&
{1 \over \beta}\sum_{{\sib p},\omega_m}
{\rm Tr}
\Bigl[
\tau_i
{\hat G}({\bf p}+{{\bf q} \over 2},i\omega_m+i\nu_n)
\tau_j
{\hat G}({\bf p}-{{\bf q} \over 2},i\omega_m)
\Bigr],
\label{eq.2.21}
\end{eqnarray}
where the second line ($\equiv\Pi_{ij}^0$) is the approximation neglecting the effect of the interactions $-U$ and $g_{\rm r}$. 
\par
Eq. (\ref{eq.2.21}) also defines other correlation functions, which will be important, such as $\Pi_{11}$ and $\Pi_{12}$. Physically, $\Pi_{11}$ and $\Pi_{22}$ describe, respectively, the amplitude and phase fluctuations of Cooper-pairs. $\Pi_{33}$ describes density fluctuations in the gas of Fermi atoms. $\Pi_{ij}$ with $i\ne j$ describes a coupling between fluctuations, e.g., $\Pi_{12}$ is a coupling of amplitude fluctuations with the phase fluctuations (amplitude-phase coupling). 
\par
Summing up the Matsubara frequencies in Eq. (\ref{eq.2.21}), we obtain\cite{OhashiT1,WongT}
\begin{eqnarray}
\Pi_{11}^0
&=&
\sum_{\sib p}
\Bigl(
1-
{
\xi_{{\sib p}+{{\sib q} \over 2}}\xi_{{\sib p}-{{\sib q} \over 2}}-{\tilde \Delta}^2
\over
E_{{\sib p}+{{\sib q} \over 2}}E_{{\sib p}-{{\sib q} \over 2}}
}
\Bigr)
{
E_{{\sib p}+{{\sib q} \over 2}}-E_{{\sib p}-{{\sib q} \over 2}}
\over
(E_{{\sib p}+{{\sib q} \over 2}}-E_{{\sib p}-{{\sib q} \over 2}})^2+\nu_n^2
}
\Bigl[
f(E_{{\sib p}+{{\sib q} \over 2}})-f(E_{{\sib p}-{{\sib q} \over 2}})
\Bigr]
\nonumber
\\
&-&
\sum_{\sib p}
\Bigl(
1+
{
\xi_{{\sib p}+{{\sib q} \over 2}}\xi_{{\sib p}-{{\sib q} \over 2}}-{\tilde \Delta}^2
\over
E_{{\sib p}+{{\sib q} \over 2}}E_{{\sib p}-{{\sib q} \over 2}}
}
\Bigr)
{
E_{{\sib p}+{{\sib q} \over 2}}+E_{{\sib p}-{{\sib q} \over 2}}
\over
(E_{{\sib p}+{{\sib q} \over 2}}+E_{{\sib p}-{{\sib q} \over 2}})^2+\nu_n^2
}
\Bigl[
1-f(E_{{\sib p}+{{\sib q} \over 2}})-f(E_{{\sib p}-{{\sib q} \over 2}})
\Bigr],
\label{eq.2.21a}
\end{eqnarray}
\begin{eqnarray}
\Pi_{22}^0
&=&
\sum_{\sib p}
\Bigl(
1-
{
\xi_{{\sib p}+{{\sib q} \over 2}}\xi_{{\sib p}-{{\sib q} \over 2}}+{\tilde \Delta}^2
\over
E_{{\sib p}+{{\sib q} \over 2}}E_{{\sib p}-{{\sib q} \over 2}}
}
\Bigr)
{
E_{{\sib p}+{{\sib q} \over 2}}-E_{{\sib p}-{{\sib q} \over 2}}
\over
(E_{{\sib p}+{{\sib q} \over 2}}-E_{{\sib p}-{{\sib q} \over 2}})^2+\nu_n^2
}
\Bigl[
f(E_{{\sib p}+{{\sib q} \over 2}})-f(E_{{\sib p}-{{\sib q} \over 2}})
\Bigr]
\nonumber
\\
&-&
\sum_{\sib p}
\Bigl(
1+
{
\xi_{{\sib p}+{{\sib q} \over 2}}\xi_{{\sib p}-{{\sib q} \over 2}}+{\tilde \Delta}^2
\over
E_{{\sib p}+{{\sib q} \over 2}}E_{{\sib p}-{{\sib q} \over 2}}
}
\Bigr)
{
E_{{\sib p}+{{\sib q} \over 2}}+E_{{\sib p}-{{\sib q} \over 2}}
\over
(E_{{\sib p}+{{\sib q} \over 2}}+E_{{\sib p}-{{\sib q} \over 2}})^2+\nu_n^2
}
\Bigl[
1-f(E_{{\sib p}+{{\sib q} \over 2}})-f(E_{{\sib p}-{{\sib q} \over 2}})
\Bigr],
\label{eq.2.21b}
\end{eqnarray}
\begin{eqnarray}
\Pi_{33}^0
&=&
\sum_{\sib p}
\Bigl(
1+
{
\xi_{{\sib p}+{{\sib q} \over 2}}\xi_{{\sib p}-{{\sib q} \over 2}}-{\tilde \Delta}^2
\over
E_{{\sib p}+{{\sib q} \over 2}}E_{{\sib p}-{{\sib q} \over 2}}
}
\Bigr)
{
E_{{\sib p}+{{\sib q} \over 2}}-E_{{\sib p}-{{\sib q} \over 2}}
\over
(E_{{\sib p}+{{\sib q} \over 2}}-E_{{\sib p}-{{\sib q} \over 2}})^2+\nu_n^2
}
\Bigl[
f(E_{{\sib p}+{{\sib q} \over 2}})-f(E_{{\sib p}-{{\sib q} \over 2}})
\Bigr]
\nonumber
\\
&-&
\sum_{\sib p}
\Bigl(
1-
{
\xi_{{\sib p}+{{\sib q} \over 2}}\xi_{{\sib p}-{{\sib q} \over 2}}-{\tilde \Delta}^2
\over
E_{{\sib p}+{{\sib q} \over 2}}E_{{\sib p}-{{\sib q} \over 2}}
}
\Bigr)
{
E_{{\sib p}+{{\sib q} \over 2}}+E_{{\sib p}-{{\sib q} \over 2}}
\over
(E_{{\sib p}+{{\sib q} \over 2}}+E_{{\sib p}-{{\sib q} \over 2}})^2+\nu_n^2
}
\Bigl[
1-f(E_{{\sib p}+{{\sib q} \over 2}})-f(E_{{\sib p}-{{\sib q} \over 2}})
\Bigr].
\label{eq.2.21b3}
\end{eqnarray}
The correlation functions $\Pi_{ij}^0$ ($i\ne j$) describing the coupling of different operators are given by
\begin{eqnarray}
\Pi_{12}^0
&=&
\sum_{\sib p}
\Bigl(
{\xi_{{\sib p}+{{\sib q} \over 2}} \over E_{{\sib p}+{{\sib q} \over 2}}}
-
{\xi_{{\sib p}-{{\sib q} \over 2}} \over E_{{\sib p}-{{\sib q} \over 2}}}
\Bigr)
{
\nu_n
\over
(E_{{\sib p}+{{\sib q} \over 2}}-E_{{\sib p}-{{\sib q} \over 2}})^2+\nu_n^2
}
\Bigl[
f(E_{{\sib p}+{{\sib q} \over 2}})-f(E_{{\sib p}-{{\sib q} \over 2}})
\Bigr]
\nonumber
\\
&-&
\sum_{\sib p}
\Bigl(
{\xi_{{\sib p}+{{\sib q} \over 2}} \over E_{{\sib p}+{{\sib q} \over 2}}}
+
{\xi_{{\sib p}-{{\sib q} \over 2}} \over E_{{\sib p}-{{\sib q} \over 2}}}
\Bigr)
{
\nu_n
\over
(E_{{\sib p}+{{\sib q} \over 2}}+E_{{\sib p}-{{\sib q} \over 2}})^2+\nu_n^2
}
\Bigl[
1-f(E_{{\sib p}+{{\sib q} \over 2}})-f(E_{{\sib p}-{{\sib q} \over 2}})
\Bigr],
\label{eq.2.21c}
\end{eqnarray}
\begin{eqnarray}
\Pi_{23}^0
&=&
-{\tilde \Delta}\nu_n
\sum_{\sib p}
\Bigl(
{1 \over E_{{\sib p}+{\sib q}/2}}
-
{1 \over E_{{\sib p}-{\sib q}/2}}
\Bigr)
{
1
\over
(E_{{\sib p}+{\sib q}/2}-E_{{\sib p}-{\sib q}/2})^2+\nu_n^2
}
\Bigl[
f(E_{{\sib p}+{\sib q}/2})-f(E_{{\sib p}-{\sib q}/2})
\Bigr]
\nonumber
\\
&+&
{\tilde \Delta}\nu_n
\sum_{\sib p}
\Bigl(
{1 \over E_{{\sib p}+{\sib q}/2}}
+
{1 \over E_{{\sib p}-{\sib q}/2}}
\Bigr)
{
1
\over
(E_{{\sib p}+{\sib q}/2}+E_{{\sib p}-{\sib q}/2})^2+\nu_n^2
}
\Bigl[
1-f(E_{{\sib p}+{\sib q}/2})-f(E_{{\sib p}-{\sib q}/2})
\Bigr],
\label{eq.5.1}
\end{eqnarray}
\begin{eqnarray}
\Pi_{13}^0
&=&
-{\tilde \Delta}
\sum_{\sib p}
{\xi_{{\sib p}+{\sib q}/2}+\xi_{{\sib p}-{\sib q}/2} 
\over 
E_{{\sib p}+{\sib q}/2} E_{{\sib p}-{\sib q}/2}
}
{
E_{{\sib p}+{\sib q}/2}-E_{{\sib p}-{\sib q}/2}
\over
(E_{{\sib p}+{\sib q}/2}-E_{{\sib p}-{\sib q}/2})^2+\nu_n^2
}
\Bigl[
f(E_{{\sib p}+{\sib q}/2})-f(E_{{\sib p}-{\sib q}/2})
\Bigr]
\nonumber
\\
&-&
{\tilde \Delta}
\sum_{\sib p}
\sum_{\sib p}
{\xi_{{\sib p}+{\sib q}/2}+\xi_{{\sib p}-{\sib q}/2} 
\over 
E_{{\sib p}+{\sib q}/2} E_{{\sib p}-{\sib q}/2}
}
{
E_{{\sib p}+{\sib q}/2}+E_{{\sib p}-{\sib q}/2}
\over
(E_{{\sib p}+{\sib q}/2}+E_{{\sib p}-{\sib q}/2})^2+\nu_n^2
}
\Bigl[
1-f(E_{{\sib p}+{\sib q}/2})-f(E_{{\sib p}-{\sib q}/2})
\Bigr],
\label{eq.5.2}
\end{eqnarray}
with $\Pi_{21}^0=-\Pi_{12}^0$, $\Pi^0_{32}=-\Pi^0_{23}$ and $\Pi^0_{31}=\Pi^0_{13}$. The density-density correlation function $\Pi_{33}^0$ and the related coupling correlation functions $\Pi_{13}^0$ and $\Pi_{23}^0$ will be used in Section VII, where we show that the Goldstone mode describing the collective phase oscillation of Cooper-pairs has spectral weight in the density correlation function. This weight comes from the coupling to the amplitude-density ($\Pi_{13}^0$) and phase-density ($\Pi_{23}^0$) correlation. 
\par
The factor $E_{{\sib p}+{{\sib q} \over 2}}-E_{{\sib p}-{{\sib q} \over 2}}$ in the denominator of the first line in Eqs. (\ref{eq.2.21a})-(\ref{eq.5.2}) describes scattering between excitations with momenta ${\bf p}\pm{\bf q}/2$ in the {\it same} quasi-particle band $E_{\sib p}$. For this reason, the first line in Eqs. (\ref{eq.2.21a})-(\ref{eq.5.2}) is referred to as the {\it intraband} term\cite{OhashiT1}. Since the thermal excitations of fermion quasi-particles are absent at $T=0$, the intraband term vanishes at $T=0$. On the other hand, the second line in Eqs. (\ref{eq.2.21a})-(\ref{eq.5.2}) is finite even at $T=0$. The factor $E_{{\sib p}+{{\sib q} \over 2}}+E_{{\sib p}-{{\sib q} \over 2}}$ in the denominator describes {\it interband} scattering between $E_{{\sib p}+{{\sib q} \over 2}}$ and $-E_{{\sib p}-{{\sib q} \over 2}}$, and thus the second line in Eqs. (\ref{eq.2.21a})-(\ref{eq.5.2}) is called the {\it interband} term. The intraband term is known to give rise to Landau damping of collective modes {\it below} the excitation gap $2{\tilde \Delta}$, while the damping due to the interband term only exists {\it above} $2{\tilde \Delta}$\cite{OhashiT1}. 
We note that the gap equation in Eq. (\ref{eq.2.17}) can be neatly expressed in terms of $\Pi_{22}^0$ in Eq. (\ref{eq.2.21b}), namely
\begin{equation}
1+{U_{\rm eff} \over 2}\Pi_{22}^0(0,0)=0.
\label{eq.2.21d}
\end{equation}
\par
The fluctuation contribution ($\delta\Omega_1$) involving the amplitude fluctuations of Cooper-pairs is similarly obtained by summing up the loop-type diagrams in Fig. 4(a) associated with the amplitude-amplitude interaction $-{U \over 4}\sum_{\sib q}\rho_{1{\sib q}}\rho_{1-{\sib q}}$. This gives
\begin{equation}
\delta\Omega_1={1 \over 2\beta}\sum_{{\sib q},\nu_n}
\ln
\Bigl[
1+{U \over 2}\Pi^0_{11}({\bf q},i\nu_n)
\Bigr].
\label{eq.2.22}
\end{equation}
The phase and amplitude fluctuations are coupled with each other through the amplitude-phase coupling $\Pi^0_{12}$ in Eq. (\ref{eq.2.21c}). This additional coupling effect can be formally incorporated into $\delta\Omega_1$ by replacing $\Pi_{11}^0({\bf q},i\nu_n)$ in Eq. (\ref{eq.2.22}) by ${\bar \Pi}_{11}({\bf q},i\nu_n)$, where 
\begin{equation}
{\bar \Pi}_{11}({\bf q},i\nu_n)\equiv\Pi_{11}^0({\bf q},i\nu_n)+\Pi_{12}^0({\bf q},i\nu_n){-U/2 \over 1+(U/2)\Pi^0_{22}({\bf q},i\nu_n)}\Pi_{21}^0({\bf q},i\nu_n).
\label{eq.2.23}
\end{equation}
The second term describes the amplitude-phase coupling effect through the coupling correlation functions $\Pi^0_{12}$ and $\Pi^0_{21}$. Eq. (\ref{eq.2.23}) is obtained by summing up the fluctuation diagrams shown in Fig. 4(b). 
\par
In summary, the fluctuation contribution $\delta\Omega_U$ to the thermodynamic potential involving only the non-resonant interaction $-U$ is the sum of Eqs. (\ref{eq.2.20}) and (\ref{eq.2.23}) with $\Pi^0_{11}\to{\bar \Pi}_{11}$. The sum is given by
\begin{eqnarray}
\delta\Omega_U
&=&
{1 \over 2\beta}
\sum_{{\sib q},\nu_n}
\ln
\Bigl[
\bigl[
1+{U \over 2}{\bar \Pi}_{11}({\bf q},i\nu_n)
\bigr]
\bigl[
1+{U \over 2}\Pi^0_{22}({\bf q},i\nu_n)
\bigr]
\Bigr]
\nonumber
\\
&=&
{1 \over 2\beta}
\sum_{{\sib q},\nu_n}
\ln
\Bigl[
\bigl[
1+{U \over 2}\Pi^0_{11}({\bf q},i\nu_n)
\bigr]
\bigl[
1+{U \over 2}\Pi^0_{22}({\bf q},i\nu_n)
\bigr]
-
\bigl({U \over 2}\bigr)^2\Pi_{12}^0({\bf q},i\nu_n)\Pi_{21}^0({\bf q},i\nu_n)
\Bigr]
\nonumber
\\
&=&
{1 \over 2\beta}\sum_{{\sib q},\nu_n}
\ln{\rm det}
\Bigl[
1+{U \over 2}{\hat \Pi}^0({\bf q},i\nu_n)
\Bigr]
\nonumber
\\
&=&
{1 \over 2\beta}\sum_{{\sib q},\nu_n}
{\rm Tr}
\ln
\Bigl[
1+{U \over 2}{\hat \Pi}^0({\bf q},i\nu_n)
\Bigr],
\label{eq.2.24}
\end{eqnarray}
where we have used the well known identity $\det{\hat A}=e^{{\rm Tr}[\ln{\hat A}]}$ in the last expression. ${\hat \Pi}^0({\bf q},i\nu_n)$ is a $2\times 2$-matrix density correlation function, defined by
\begin{eqnarray}
{\hat \Pi}^0({\bf q},i\nu_n)=
\left(
\begin{array}{cc}
\Pi_{11}^0 & \Pi_{12}^0 \\
\Pi_{21}^0 & \Pi_{22}^0
\end{array}
\right).
\label{eq.2.25}
\end{eqnarray}
\par
Finally we consider the fluctuation contributions $\delta\Omega_{\rm FR}$ from the Feshbach resonance term ${1 \over 2}g_{\rm r}\sum_{\sib q}[b^\dagger_{\sib q}\rho_{\sib q}^-+b_{\sib q}\rho_{\sib q}^+]$ in Eq. (\ref{eq.2.13}). First we sum up the diagrams described in Fig. 4(a), where the dashed line now represents the $b$-boson Green's function $D_0({\bf q},i\nu_n)$ in Eq. (\ref{eq.2.10}), to give
\begin{equation}
\delta\Omega_{\rm FR}=
{1 \over 2\beta}
\sum_{{\sib q},\nu_n}{\rm Tr}
\ln
\Bigl[
1-{g_{\rm r}^2 \over 4}{\hat D}^0({\bf q},i\nu_n){\hat \Xi}^0({\bf q},i\nu_n)
\Bigr].
\label{eq.2.26}
\end{equation}
Here ${\hat D}^0({\bf q},i\nu_n)$ is a $2\times 2$-matrix $b$-boson Green's function defined by
\begin{eqnarray}
{\hat D}^0({\bf q},\nu_n)=
{1 \over i\nu_n\tau_3-\xi_{B\sib q}}
=
-
{
i\nu_n\tau_3+\xi_{B{\sib q}}
\over 
\nu_n^2+\xi_{B{\sib q}}^2
}
,
\label{eq.2.27}
\end{eqnarray}
and ${\hat \Xi}^0({\bf q},i\nu_n)$ is a $2\times 2$-matrix correlation function (neglecting the effect of $-U$ and $g_{\rm r}$) defined by
\begin{eqnarray}
{\hat \Xi}^0({\bf q},i\nu_n)=
-\int_0^\beta d\tau e^{i\nu_n\tau}
\Bigl\langle
T_{\tau}
\Bigl[
\left(
\begin{array}{cc}
\rho^-_{\sib q}(\tau)
\rho^+_{-\sib q}(0) 
&
\rho^-_{\sib q}(\tau)
\rho^-_{-\sib q}(0) 
\\
\rho^+_{\sib q}(\tau)
\rho^+_{-\sib q}(0) 
&
\rho^+_{\sib q}(\tau)
\rho^-_{-\sib q}(0) 
\end{array}
\right)
\Bigr]
\Bigr\rangle
.
\label{eq.2.28}
\end{eqnarray}
Using the definition $\rho^{\pm}_{\sib q}=\rho_{1\pm{\sib q}}\pm i\rho_{2\pm{\sib q}}$, we find that ${\hat \Xi}^0$ can be expressed in terms of the matrix elements of ${\hat \Pi}^0$ as follows: 
\begin{eqnarray}
{\hat \Xi}^0
&=&
2{\hat W}^{-1}{\hat \Pi}^0{\hat W}
\nonumber
\\
&=&
\left(
\begin{array}{cc}
\Pi_{11}^0+\Pi_{22}^0+i(\Pi_{12}^0-\Pi_{21}^0) &
\Pi_{11}^0-\Pi_{22}^0 \\
\Pi_{11}^0-\Pi_{22}^0 &
\Pi_{11}^0+\Pi_{22}^0-i(\Pi_{12}^0-\Pi_{21}^0) 
\end{array}
\right),
\label{eq.2.29}
\end{eqnarray}
where ${\hat W}$ is the unitary matrix 
\begin{eqnarray}
{\hat W}={1 \over \sqrt{2}}
\left(
\begin{array}{cc}
1 & 1\\
i & -i
\end{array}
\right).
\label{eq.2.30}
\end{eqnarray}
\par
Next we renormalize the fluctuations in Eq. (\ref{eq.2.26}) by including the effects of the non-resonant interaction $-U$ on the correlation function ${\hat \Xi}^0({\bf q},i\nu_n)$, working within the HF-RPA. This results in ${\hat \Xi}^0({\bf q},i\nu_n)$ in Eq. (\ref{eq.2.26}) being replaced by ${\hat \Xi}_U$, where\cite{OhashiT1,OhashiT2}
\begin{equation}
{\hat \Xi}_U({\bf q},i\nu_n)=
\Bigl[
1+{U \over 4}{\hat \Xi}^0({\bf q},i\nu_n)
\Bigr]^{-1}{\hat \Xi}^0({\bf q},i\nu_n).
\label{eq.2.31b}
\end{equation}
The resulting expression for $\delta\Omega_{\rm FR}$ involves the fluctuation effects related to both the non-resonant interaction $-U$ and the Feshbach resonance coupling parameter $g_{\rm r}$, namely
\begin{eqnarray}
\delta\Omega_{\rm FR}
&=&
{1 \over 2\beta}
\sum_{{\sib q},\nu_n}
{\rm Tr}\ln
\Bigl[
1-{g_{\rm r}^2 \over 4}{\hat D}^0({\bf q},i\nu_n)
{\hat \Xi}_U({\bf q},i\nu_n)
\Bigr]
\nonumber
\\
&=&
{1 \over 2\beta}
\sum_{{\sib q},\nu_n}
{\rm Tr}\ln
\Bigl\{
1-{g_{\rm r}^2 \over 4}{\hat D}^0({\bf q},i\nu_n)
\Bigl[
1+{U \over 4}{\hat \Xi}^0({\bf q},i\nu_n)
\Bigr]^{-1}
{\hat \Xi}^0({\bf q},i\nu_n)
\Bigr\}
.
\label{eq.2.32}
\end{eqnarray}
\par
The total fluctuation contribution $\delta\Omega$ is given by the sum of Eqs. (\ref{eq.2.24}) and (\ref{eq.2.32}). Recalling the definition in Eq.(\ref{eq.2.29}) and the relation ${\rm Tr}\ln[1+{U \over 2}{\hat \Pi}^0]={\rm Tr}\ln[{\hat W}\{1+{U \over 4}{\hat \Xi}^0\}{\hat W}^{-1}]={\rm Tr}\ln[1+{U \over 4}{\hat \Xi}^0]$, this sum can be written as
\begin{eqnarray}
\delta\Omega
&\equiv&
\delta\Omega_U+\delta\Omega_{\rm FR}
\nonumber
\\
&=&
{1 \over 2\beta}\sum_{{\sib q},\nu_n}
{\rm Tr}\ln
\Bigl\{
1+{1 \over 4}
\Bigl[
U-g_{\rm r}^2{\hat D}^0({\bf q},i\nu_n)
\Bigr]{\hat \Xi}^0({\bf q},i\nu_n)
\Bigr\}.
\label{eq.2.33}
\end{eqnarray}
Putting everything together, the total thermodynamic potential $\Omega=\Omega_{\rm MF}+\delta\Omega$ is 
\begin{eqnarray}
\Omega
&=&
{{\tilde \Delta}^2 \over U_{\rm eff}}
+\sum_{\sib p}
(\xi_{\sib p}-E_{\sib p})
-2T
\sum_{\sib p}\ln\Bigl[1+e^{-\beta E_{\sib p}}\Bigr]
+T\sum_{\sib q}\ln\Bigl[1-e^{-\beta\xi_{\sib q}^B}\Bigr]
\nonumber
\\
&+&
{1 \over 2\beta}\sum_{{\sib q},\nu_n}
{\rm Tr}\ln
\Bigl\{
1+{1 \over 4}
\Bigl[
U-g_{\rm r}^2{\hat D}^0({\bf q},i\nu_n)
\Bigr]{\hat \Xi}^0({\bf q},i\nu_n)
\Bigr\}.
\label{eq.2.34}
\end{eqnarray}
The equation for the total number of particles is then obtained from $N=-\partial\Omega/\partial\mu$. 
\par
In taking the derivative with respect to $\mu$, we note that the order parameter ${\tilde \Delta}$ also depends on the chemical potential $\mu$, in addition to $\xi_{\sib p}\equiv\varepsilon_{\sib p}-\mu$ and $\xi_{B\sib q}\equiv\varepsilon_{B\sib q}+2\nu-2\mu$. Thus one needs to calculate 
\begin{eqnarray}
{\partial \Omega \over \partial{\tilde \Delta}}
{\partial {\tilde \Delta} \over \partial \mu}
&=&
{\partial \Omega_{\rm MF} \over \partial{\tilde \Delta}}
{\partial {\tilde \Delta} \over \partial \mu}
+
{\partial \delta\Omega \over \partial{\tilde \Delta}}
{\partial {\tilde \Delta} \over \partial \mu}
\nonumber
\\
&=&
{\partial \delta\Omega \over \partial{\tilde \Delta}}
{\partial {\tilde \Delta} \over \partial \mu}.
\label{eq.2.35}
\end{eqnarray}
In the last line, we have used the fact $\partial\Omega_{\rm MF}/\partial{\tilde \Delta}=0$, which holds when ${\tilde \Delta}$ satisfies the gap equation (\ref{eq.2.17}). However, as will be shown in Section IV, since $\partial\delta\Omega/\partial{\tilde \Delta}$ in Eq. (\ref{eq.2.35}) is a higher order correction within the perturbative approximation we are using, we can neglect the contribution from Eq. (\ref{eq.2.35}). Thus the dependence of ${\tilde \Delta}$ on $\mu$ only leads to higher order corrections. The resulting equation for $N$ is 
\begin{eqnarray}
N
&=&
2\phi_{\rm m}^2
+\sum_{\sib p}
\Bigl[1-{\xi_{\sib p} \over E_{\sib p}}\tanh{\beta \over 2}E_{\sib p}
\Bigr]
+2\sum_{\sib q}n_{\rm B}(\xi^B_{\sib q})
\nonumber
\\
&-&
{1 \over 2\beta}\sum_{{\sib q},\nu_n}
{\partial \over \partial \mu}
{\rm Tr}\ln
\Bigl\{
1+{1 \over 4}
\Bigl[
U-g_{\rm r}^2{\hat D}^0({\bf q},i\nu_n)
\Bigr]{\hat \Xi}^0({\bf q},i\nu_n)
\Bigr\}.
\label{eq.2.36}
\end{eqnarray}
Here it is understood that the $\mu$-derivative in the last term only acts on the chemical potential involved in $\xi_{\sib p}$ and $\xi_{B\sib q}$. Eqs. (\ref{eq.2.17}) and (\ref{eq.2.36}) give us the required self-consistent coupled equations for ${\tilde \Delta}$ and $\mu$ in the superfluid phase below $T_{\rm c}$. We will discuss our numerical self-consistent solutions of the coupled equations (\ref{eq.2.17}) and (\ref{eq.2.36}) in Section VI.
\par
When one neglects the fluctuation contribution given in the last term in Eq. (\ref{eq.2.36}), we obtain the mean-field expression obtained in Ref.\cite{Ranninger2} in the context of high-$T_{\rm c}$ superconductivity. We also mention that Eq. (\ref{eq.2.36}) reproduces Eq. (\ref{eq.2.9}) for the normal phase $T\ge T_{\rm c}$, where $\Delta=\phi_{\rm m}=0$. To see this, we note that the second term on the right hand side in Eq. (\ref{eq.2.36}) reduces to $2N_{\rm F}^0$ in Eq. (\ref{eq.2.9}). In addition, since the phase and amplitude fluctuations are indistinguishable when $\Delta=0$, we find $\Pi_{11}^0=\Pi_{22}^0$, and thus ${\hat \Xi}_0$ in Eq. (\ref{eq.2.29}) becomes diagonal at $T_{\rm c}$. Noting that $\Xi^0_{11}({\bf q},i\nu_n,T=T_{\rm c})=-4\Pi({\bf q},i\nu_n)$ and $\Xi^0_{22}({\bf q},i\nu_n,T=T_{\rm c})=-4\Pi({\bf q},-i\nu_n)$, where $\Pi({\bf q},i\nu_n)$ is given in Eq. (\ref{eq.2.11}), we find that the last term in Eq. (\ref{eq.2.36}) ($\equiv\delta N$) reproduces the last term in Eq. (\ref{eq.2.9}). More explicitly, we have 
\begin{eqnarray}
\delta N(T_{\rm c})
&=&
-
{1 \over 2\beta}\sum_{{\sib q},\nu_n}
{\partial \over \partial \mu}
\ln
\Bigl[
\Bigl(
1-[U-g_{\rm r}^2D^0_{11}({\bf q},i\nu_n)]\Pi({\bf q},i\nu_n)
\Bigr)
\nonumber
\\
&{ }&~~~~~~~~~~~~~~~~~~~\times
\Bigl(
1-[U-g_{\rm r}^2D^0_{22}({\bf q},i\nu_n)]\Pi({\bf q},-i\nu_n)
\Bigr)
\Bigr]
\nonumber
\\
&=&
-{1 \over \beta}\sum_{{\sib q},\nu_n}
{\partial \over \partial \mu}
\ln
\Bigl\{
1-[U-g_{\rm r}^2D_0({\bf q},i\nu_n)]\Pi({\bf q},i\nu_n)
\Bigr\},
\label{eq.2.37}
\end{eqnarray}
where we have used $D^0_{11}({\bf q},i\nu_n)=D_0({\bf q},i\nu_n)$ and $D^0_{22}({\bf q},i\nu_n)=D_0({\bf q},-i\nu_n)$ in the last line. Thus Eq. (\ref{eq.2.36}) is equivalent to Eq. (\ref{eq.2.9}) at $T_{\rm c}$. Our present strong-coupling theory giving the coupled equations (\ref{eq.2.17}) and (\ref{eq.2.36}) for the superfluid phase is seen to smoothly go over to our previous discussion at $T_{\rm c}$ and above\cite{Ohashi1,Ohashi2,Ohashi3}.
\par
Each term in Eq. (\ref{eq.2.36}) has a simple physical meaning, which is useful to discuss. The first term 
\begin{equation}
2N_{\rm B}^{\rm c}\equiv 2\phi_{\rm m}^2=2\langle b_{{\sib q}=0}\rangle^2
\label{eq.2.37a}
\end{equation}
gives twice the number of Bose-{\it condensed} $b$-bosons. The second term 
\begin{equation}
N_{\rm F}\equiv\sum_{\sib p}[1-{\xi_{\sib p} \over E_{\sib p}}\tanh{\beta \over 2}E_{\sib p}]
\label{eq.2.37b}
\end{equation}
describes the number of Fermi quasi-particles. This expression can be directly obtained from $N_{\rm F}=\sum_{{\sib p},\sigma}\langle c_{{\sib p}\sigma}^\dagger c_{{\sib p}\sigma}\rangle$ in Eq. (\ref{eq.2.2}). 
To understand the physical meanings of the last two terms in Eq. (\ref{eq.2.36}), it is convenient to divide the $\mu$-derivative in the last term into the derivative acting on $\xi_{B\sib q}=q^2/2M+2\nu-2\mu$ in the $b$-boson Green's function ${\hat D}^0({\bf q},i\nu_n)$ and the derivative acting on $\xi_{\sib p}=\varepsilon_{\sib p}-\mu$ in $\Xi^0({\bf q},i\nu_n)$ ($\equiv \partial/\partial\mu_{\rm F}$). Using the identity
\begin{equation}
2\sum_{\sib q}n_{\rm B}(\xi_{B\sib q})=
-\sum_{\sib q}
\Bigl[1+{1 \over \beta}\sum_{\nu_n}
{\rm Tr}[{\hat D}^0({\bf q},i\nu_n)]
\Bigr],
\label{eq.2.59}
\end{equation}
we can write Eq. (\ref{eq.2.36}) as
\begin{equation}
N=2N_{\rm B}^{\rm c}+N_{\rm F}+2N_{\rm B}^{\rm n}+2N_{\rm C},
\label{eq.2.59b}
\end{equation}
where $N_{\rm B}^{\rm n}$ and $N_{\rm C}$ are defined by
\begin{eqnarray}
\left\{
\begin{array}{l}
\displaystyle
2N^{\rm n}_{\rm B}\equiv
-\sum_{\sib q}
\Bigl[1+{1 \over \beta}\sum_{\nu_n}
{\rm Tr}[{\hat D}({\bf q},i\nu_n)]
\Bigr],
\\
\displaystyle
2N_{\rm C}\equiv
-
{1 \over 2\beta}\sum_{{\sib q},\nu_n}
{\partial \over \partial \mu_{\rm F}}
{\rm Tr}\ln
\Bigl\{
1+{1 \over 4}
\Bigl[
U-g_{\rm r}^2{\hat D}^0({\bf q},i\nu_n)
\Bigr]{\hat \Xi}^0({\bf q},i\nu_n)
\Bigr\}.
\end{array}
\right.
\label{eq.2.60}
\end{eqnarray}
Here ${\hat D}({\bf q},i\nu_n)$ is a renormalized $2\times 2$-matrix thermal $b$-boson Green's function defined by
\begin{equation}
{\hat D}({\bf q},i\nu_n)=
{1
\over 
i\nu_n\tau_3-\xi_{B\sib q}-{\hat \Sigma}({\bf q},i\nu_n)
}.
\label{eq.2.61}
\end{equation}
The $b$-boson matrix self-energy
\begin{equation}
{\hat \Sigma}({\bf q},i\nu_n)
\equiv{g_{\rm r}^2 \over 4}{\hat \Xi}_U({\bf q},i\nu_n)
={g_{\rm r}^2 \over 4}
\Bigl[1+{U \over 4}{\hat \Xi}^0({\bf q},i\nu_n)\Bigr]^{-1}
{\hat \Xi}^0({\bf q},i\nu_n),
\label{eq.2.62}
\end{equation}
describes the Feshbach coupling with Fermi atoms. 
\par
The renormalized $b$-boson Green's function in Eq. (\ref{eq.2.61}) has the same form as that obtained by Kostyrko and Ranninger\cite{Ranninger2} calculated within the HF-RPA. Comparing $N^{\rm n}_{\rm B}$ in Eq. (\ref{eq.2.60}) with Eq. (\ref{eq.2.59}), we may interpret $N^{\rm n}_{\rm B}$ as the number of {\it non-condensed} $b$-bosons, as renormalized by the Feshbach resonance. In analogy to our previous discussion of the strong-coupling theory for $T_{\rm c}$\cite{Ohashi1,Ohashi2,Ohashi3}, $N_{\rm C}$ in Eq. (\ref{eq.2.60}) may be understood as the fluctuation contribution to $N$ from Cooper-pairs associated with the (dynamical) pairing interaction given by
\begin{equation}
{\hat U}_{\rm eff}({\bf q},i\nu_n)\equiv
U-g_{\rm r}^2{\hat D}^0({\bf q},i\nu_n).
\label{eq.2.63}
\end{equation}
We note that ${\hat U}_{\rm eff}({\bf q}=0,i\nu_n=0)=U_{\rm eff}{\hat {\bf 1}}$, where $U_{\rm eff}=U+g_{\rm r}^2/(2\nu-2\mu)$ is the effective pairing interaction appearing in the gap equation in Eq. (\ref{eq.2.17}).
\par
The renormalized $b$-boson Green's function in Eq. (\ref{eq.2.61}) can be also written in the form
\begin{eqnarray}
{\hat D}({\bf q},i\nu_n)
&=&
{1
\over 
i{\tilde \nu}_n\tau_3-{\tilde \xi}_{B\sib q}-\kappa\tau_1
}
=
-{i{\tilde \nu}_n\tau_3+{\tilde \xi}_{B\sib q}-\kappa\tau_1
\over
{\tilde \nu}_n^2+{{\tilde \xi}^{2}_{B\sib q}}-\kappa^2
}
,
\label{eq.2.64}
\end{eqnarray}
where the renormalized parameters are given by
\begin{eqnarray}
\left\{
\begin{array}{l}
\displaystyle
i{\tilde \nu}_n\equiv i\nu_n-i{g_{\rm r}^2 \over 4}[\Pi_{U12}-\Pi_{U21}],
\\
\displaystyle
{\tilde \xi}_{B\sib q}\equiv \xi_{B\sib q}
+{g_{\rm r}^2 \over 4}[\Pi_{U11}+\Pi_{U22}],
\\
\displaystyle
\kappa\equiv{g_{\rm r}^2 \over 4}[\Pi_{U11}-\Pi_{U22}].
\end{array}
\right.
\label{eq.2.65}
\end{eqnarray}
The correlation functions $\Pi_{Uij}$ ($i,j=1,2$) are the matrix elements of ${\hat \Pi}_U$ defined by
\begin{equation}
{\hat \Pi}_U({\bf q},i\nu_n)=
\Bigl[
1+{U \over 2}{\hat \Pi}^0({\bf q},i\nu_n)
\Bigr]^{-1}{\hat \Pi}^0({\bf q},i\nu_n).
\label{eq.2.31}
\end{equation}
More explicitly, we have
\begin{eqnarray}
\left\{
\begin{array}{l}
\displaystyle
\Pi_{U11}={{\bar \Pi}_{11} \over 1+{U \over 2}{\bar \Pi}_{11}},
\\
\displaystyle
\Pi_{U22}={{\bar \Pi}_{22} \over 1+{U \over 2}{\bar \Pi}_{22}},
\\
\displaystyle
\Pi_{U12}=
{\Pi^0_{12} \over 
[1+{U \over 2}\Pi^0_{11}][1+{U \over 2}\Pi^0_{22}]-{U^2 \over 4}\Pi^0_{12}\Pi^0_{21}
},
\\
\Pi_{U21}({\bf q},i\nu_n)=-\Pi_{U12}({\bf q},i\nu_n).
\end{array}
\right.
\label{eq.2.66}
\end{eqnarray}
Here ${\bar \Pi}_{11}$ is defined in Eq. (\ref{eq.2.23}), and ${\bar \Pi}_{22}$ is similarly obtained from Eq. (\ref{eq.2.23}) by interchanging $1\leftrightarrow 2$. 
\par
When we take ${\bf q}=i\nu_n=0$, the denominator of Eq. (\ref{eq.2.64}) reduces to
\begin{eqnarray}
{\tilde \xi}_{B{\sib q}=0}^2-\kappa(0,0)^2=
[1+{U_{\rm eff} \over 2}\Pi^0_{22}(0,0)]
[1+{U_{\rm eff} \over 2}\Pi^0_{11}(0,0)]
{2\nu-2\mu \over 1+{U \over 2}\Pi_{22}^0(0,0)}
{2\nu-2\mu \over 1+{U \over 2}\Pi_{11}^0(0,0)}.
\label{eq.2.67}
\end{eqnarray}
The expression in Eq. (\ref{eq.2.67}) clearly vanishes when the gap equation in Eq. (\ref{eq.2.21d}) is satisfied. This means that the excitation spectrum of the renormalized $b$-bosons described by ${\hat D}$ is always {\it gapless} at ${\bf q}=0$ for $T<T_{\rm c}$. This is a desired result, because the Bose-condensation of $b$-bosons characterized by $\phi_{\rm m}$ should be accompanied by the appearance of a Bogoliubov phonon (Goldstone) mode having a (uniform system) gapless dispersion. Thus the strong-coupling theory presented in this section correctly includes a gapless spectrum for the symmetry breaking Goldstone mode. This result is not obtained if we use a static mean-field theory, neglecting the fluctuation contribution given by the last term of Eq. (\ref{eq.2.36}). Within such a mean-field theory, the $b$-boson excitation spectrum is given by $\varepsilon_{B\sib q}+2\nu-2\mu$, which always has a {\it finite excitation gap} at ${\bf q}=0$. This gap is given by $\delta E\equiv 2\nu-2\mu=-{g_{\rm r}^2 \over 2}\Pi_{U22}(0,0)>0$, assuming that the gap equation in Eq. (\ref{eq.2.21d}) is satisfied. 
We note that $\Pi_{U22}(0,0)$ is defined in Eq. (\ref{eq.2.66}), with ${\bar \Pi}_{22}(0,0)=\Pi^0_{22}(0,0)$ since $\Pi_{12}^0(0,0)=0$.
\par
The gapless behavior of the renormalized $b$-boson can be also verified formally by noting that the Hugenholtz-Pines theorem is verified, namely\cite{Ranninger2,Pines1,Pines2,Rickayzen}
\begin{equation}
\Sigma_{11}(0,0)-\Sigma_{12}(0,0)=
{g_{\rm r}^2 \over 2}
{\Pi^0_{22}(0,0) \over 1+{U \over 2}\Pi^0_{22}(0,0)}=2\mu-2\nu.
\label{eq.2.67b}
\end{equation}
Here we have used the gap equation given in Eq. (\ref{eq.2.21d}).
\par
In order to discuss BCS-BEC crossover in the superfluid phase, one needs to solve the coupled equations (\ref{eq.2.17}) and (\ref{eq.2.36}) numerically. Results will be discussed in Section VI. Here we briefly discuss limiting cases which can be treated analytically. 
\par
When the threshold energy $2\nu$ of the $b$-boson excitation spectrum is very large (BCS limit), $2\nu\gg2\mu$ can be realized because the chemical potential $\mu$ is at most the order of $\varepsilon_{\rm F}$. In this case, the Fermi atoms are dominant particles while the effects of $b$-molecules described by $\phi_{\rm m}$, $n_{\rm B}(\xi_{B\sib q})$ and ${\hat D}^0$ in Eq. (\ref{eq.2.36}) can be neglected. In addition, since the last term in Eq. (\ref{eq.2.36}) is small when one is dealing with a weak non-resonant interaction $-U$, one can also drop this term. Thus we see that Eq. (\ref{eq.2.36}) reduces to $N=N_{\rm F}$ (where $N_{\rm F}$ is defined in Eq. (\ref{eq.2.37b})), which simply gives $\mu\simeq\varepsilon_{\rm F}$ as far as $\varepsilon_{\rm F}\gg T$. When this result is substituted into the gap equation in Eq. (\ref{eq.2.17}) with $U_{\rm eff}\to U$ and ${\tilde \Delta}\to\Delta$ (note that $2\nu\gg2\mu$.), one obtains the usual BCS gap equation for the Cooper-pair order parameter $\Delta$.
\par
In the opposite limit $2\nu\lesssim-2\varepsilon_{\rm F}$ (BEC limit), since the $b$-boson branch has an energy lower than the two fermion band energy, most Fermi atoms will combine to form $b$-molecules and hence the fermion correlation functions $\Pi_{ij}$ becomes less important. Then the gap equation in Eq. (\ref{eq.2.21d}) can be rewritten as [see also Eq. (\ref{eq.2.67b})]
\begin{equation}
2\mu=2\nu+{g_{\rm r}^2 \over 2}
{\Pi^0_{22}(0,0) \over 1+{U \over 2}\Pi^0_{22}(0,0)}
\to2\nu.
\label{eq.2.67c}
\end{equation}
This says that the chemical potential has the energy of the bottom of the $b$-boson excitation spectrum. Substituting $2\mu=2\nu$ into Eq. (\ref{eq.2.36}) with ${\hat \Xi}^0({\bf q},i\nu_n)=0$, we obtain
\begin{equation}
{N \over 2}=\phi_{\rm m}+\sum_{\sib q}n_{\rm B}({q^2 \over 2M}).
\label{eq.2.67d}
\end{equation}
This is just the equation for BEC in an non-interacting uniform Bose gas with $N/2$ bosons of mass $M$. Thus the coupled Eqs. (\ref{eq.2.17}) and (\ref{eq.2.36}) reproduce both the BCS-phase and BEC-phase for two limiting values of the Feshbach molecular resonance threshold $2\nu$.
\par
We again remind the reader about the many body approximation our whole discussion is based on. As we noted after Eq. (\ref{eq.2.17}), we have included the fluctuations of the order parameters $\Delta$ and $\phi_{\rm m}$ in determining self-consistently the values of $N$ and $\mu$. However, we have not included the explicit self-energy of the fermions due to these dynamic fluctuations. For further discussion, see Refs.\cite{Cote,Haussmann2}.
\vskip2mm
\section{Relation to a Gaussian fluctuation theory}
\vskip3mm
It is interesting see what type of fluctuations are taken into account in the strong-coupling theory presented in Section III. In this Section, we show that the coupled Eqs. (\ref{eq.2.17}) and (\ref{eq.2.36}) describe {\it Gaussian fluctuations} around the mean-field order parameters $\Delta$ and $\phi_{\rm m}$. To see this, we employ the functional integral formalism because it treats fluctuations in a very clear fashion. Indeed, using this formalism, Randeria and co-workers\cite{Melo,Engelbrecht} showed that the strong-coupling superconductivity theory developed by Nozi\`eres and Schmitt-Rink\cite{Nozieres} can be understood as a Gaussian approximation for the fluctuations in the Cooper-channel. Very recently, Milstein et al.\cite{Milstein} also employed this approach to the coupled fermion-boson model in Eq. (\ref{eq.2.1}) and reproduced the equations for $T_{\rm c}$ given in Eqs. (\ref{eq.2.7}) and (\ref{eq.2.9}). 
\par
The partition function for the coupled fermion-boson model in Eq. (\ref{eq.2.1}) in the functional integral formalism is given by
\begin{equation}
Z=\int
{\cal D}\Psi^\dagger_\sigma
{\cal D}\Psi_\sigma
{\cal D}\Phi^\dagger
{\cal D}\Phi
e^{-S}.
\label{eq.2.38}
\end{equation}
The action $S$ has the form
\begin{eqnarray}
S
&=&
\int_0^\beta d\tau\int d{\bf r}
\Bigl[
\sum_\sigma
\Psi^\dagger_\sigma(x)
\Bigl[
\partial_\tau+{{\hat {\bf p}}^2 \over 2m}-\mu
\Bigr]
\Psi_\sigma(x)
-
U
\Psi^\dagger_\uparrow(x)
\Psi^\dagger_\downarrow(x)
\Psi_\downarrow(x)
\Psi_\uparrow(x)
\nonumber
\\
&+&
\Phi^\dagger(x)
\Bigl[
\partial_\tau+{{\hat {\bf p}}^2 \over 2M}+2\nu-2\mu
\Bigr]
\Phi(x)
+
g_{\rm r}
[\Phi^\dagger(x)\Psi_\downarrow(x)\Psi_\uparrow(x)+
\Phi(x)\Psi^\dagger_\uparrow(x)\Psi^\dagger_\downarrow(x)]
\Bigr],
\label{eq.2.39}
\end{eqnarray}
where ${\hat {\bf p}}\equiv\nabla/i$ and $x\equiv({\bf r},\tau)$. $\Psi_\sigma(x)$ is a Grassmann variable for Fermi atoms and $\Psi_\sigma^\dagger(x)$ is its conjugate. $\Phi(x)$ is a complex scalar Bose field describing $b$-molecules. When we introduce the Cooper-pair field $\Delta$ using the Stratonovich-Hubbard transformation, the partition function in Eq. (\ref{eq.2.38}) is transformed to 
\begin{eqnarray}
Z=\int
{\cal D}\Psi^\dagger_\sigma
{\cal D}\Psi_\sigma
{\cal D}\Phi^\dagger
{\cal D}\Phi
{\cal D}\Delta^\dagger
{\cal D}\Delta
e^{-{\bar S}},
\label{eq.2.40a}
\end{eqnarray}
where the action ${\bar S}$ is now given by
\begin{eqnarray}
{\bar S}
&=&
\int_0^\beta d\tau\int d{\bf r}
\Bigl[
{|\Delta(x)|^2 \over U}
-{\hat \Psi}^\dagger(x){\hat G}^{-1}(x){\hat \Psi}(x)
-
{1 \over 2}{\hat \Phi}^\dagger(x){\hat D}^0(x)^{-1}{\hat \Phi}(x)
\Bigr]
\nonumber
\\
&+&
\beta\sum_{\sib p}\xi_{\sib p}
-{\beta \over 2}\sum_{\sib q}\xi_{B{\sib q}}.
\label{eq.2.40}
\end{eqnarray}
Here we have introduced the Nambu representation for fermions as ${\hat \Psi}^\dagger\equiv(\Psi_\uparrow^\dagger, \Psi_\downarrow)$. The corresponding $2\times2$-matrix thermal Green's function ${\hat G}(x)$ in the $({\bf r},\tau)$-representation is given by
\begin{eqnarray}
{\hat G}^{-1}(x)=-\partial_\tau-
\Bigl(
{{\hat {\bf p}}^2 \over 2m}-\mu
\Bigr)\tau_3
+{\hat {\tilde \Delta}}(x),
\label{eq.2.41}
\end{eqnarray}
where 
\begin{eqnarray}
{\hat {\tilde \Delta}}(x)
\equiv
\left(
\begin{array}{cc}
0& {\tilde \Delta}(x)\\
{\tilde \Delta}(x)^\dagger & 0
\end{array}
\right)
=
\left(
\begin{array}{cc}
0& \Delta(x)-g_{\rm r}\Phi(x)\\
\Delta(x)^\dagger-g_{\rm r}\Phi^\dagger(x) & 0
\end{array}
\right).
\label{eq.2.42}
\end{eqnarray}
We have also introduced a matrix notation for the $b$-bosons as ${\hat \Phi}^\dagger(x)\equiv (\Phi^\dagger(x),\Phi(x))$, as well as the corresponding 
matrix $b$-boson Green's function in the $({\bf r},\tau)$-representation,
\begin{equation}
{\hat D}^0(x)^{-1}=-\partial_\tau\tau_3-
\Bigl({{\hat {\bf p}}^2 \over 2M}+2\nu-2\mu\Bigr).
\label{eq.2.43}
\end{equation}
\par
Since the action ${\bar S}$ in Eq. (\ref{eq.2.40}) has a bi-linear form with respect to ${\hat \Psi}$ and ${\hat \Psi}^\dagger$, one can immediately integrate out the fermion degrees of freedom using the formula
\begin{equation}
\int{\cal D}\Psi_\sigma^\dagger{\cal D}\Psi_\sigma
e^{-\sum_{ij}\Psi_i^\dagger M_{ij}\Psi_j}={\rm det}[{\hat M}],
\label{eq.2.44}
\end{equation}
where $i$ and $j$ represent the variables (${\bf r},\tau,\sigma$) involved in $\Psi_\sigma({\bf r},\tau)$. After the functional integration, the action is reduced to ($\equiv {\tilde S}$)
\begin{eqnarray}
{\tilde S}
&=&
-{\rm Tr}\ln[-{\hat G}^{-1}]
+\int_0^\beta d\tau\int d{\bf r}
\Bigl[
{|\Delta(x)|^2 \over U}
-{1 \over 2}{\hat \Phi}(x)^\dagger {\hat D}^0(x)^{-1}{\hat \Phi}(x)
\Bigr]
\nonumber
\\
&+&
\beta\sum_{\sib p}\xi_{\sib p}
-{\beta \over 2}\sum_{\sib q}\xi_{B\sib q}.
\label{eq.2.45}
\end{eqnarray}
The trace in ${\rm Tr}\ln[-{\hat G}^{-1}]$ involves the integrals over $\tau$ and ${\bf r}$, as well as a trace over the particle-hole space.  Eq. (\ref{eq.2.45}) has a very physical form because it is described by two kinds of Bose fields $\Delta(x)$ and $\Phi(x)$ directly related to superfluid fluctuations.
\par
In order to extract the fluctuation contribution from Eq. (\ref{eq.2.45}), we divide the Bose fields $\Delta(x)$ and $\Phi(x)$ into their mean-field values ($\Delta$, $\phi_{\rm m}$) and fluctuations from these values. In the functional integral formalism, the mean-field approximation corresponds to the saddle point solution, given by
\begin{equation}
{\delta {\tilde S} \over \delta\Delta^\dagger(x)}=
{\delta {\tilde S} \over \delta\Phi^\dagger(x)}=0.
\label{eq.2.46}
\end{equation}
Since we are working with a uniform system (no trap), we can take a uniform (and real) solution as the saddle point solution. Then Eq. (\ref{eq.2.46}) gives the two equations:
\begin{eqnarray}
\left\{
\begin{array}{l}
\displaystyle
\Delta=U{\hat G}_{12}(x)
={U \over \beta}\sum_{{\sib p},\omega_m}{\hat G}_{12}({\bf p},i\omega_m)
={U \over \beta}
\sum_{\sib p}{{\tilde \Delta} \over 2E_{\sib p}}
\tanh{\beta \over 2}E_{\sib p},\\
\displaystyle
(2\nu-2\mu)\phi_{\rm m}=-g_{\rm r}{\hat G}_{12}(x)
=-{g_{\rm r} \over \beta}
\sum_{\sib p}{{\tilde \Delta} \over 2E_{\sib p}}
\tanh{\beta \over 2}E_{\sib p},
\end{array}
\right.
\label{eq.2.47}
\end{eqnarray}
where ${\hat G}_{12}({\bf p},i\omega_m)$ is the off-diagonal component of the fermion Green's function in Eq. (\ref{eq.2.15}). Eq. (\ref{eq.2.47}) is equivalent to Eqs. (\ref{eq.2.16})-(\ref{eq.2.18}). 
\par
Writing $\Delta(x)=\Delta+\delta\Delta(x)$ and $\Phi(x)=\phi_{\rm m}+\delta\Phi(x)$, we can express the partition function $Z$ in terms of the fluctuation fields $\delta\Delta(x)$ and $\delta\Phi(x)$ as
\begin{equation}
Z=
e^{-{\tilde S}_0}
\int{\cal D}\delta\Phi^\dagger{\cal D}\delta\Phi{\cal D}\delta\Delta^\dagger{\cal D}\delta\Delta
e^{-\delta{\tilde S}}.
\label{eq.2.48}
\end{equation}
Here the action ${\tilde S}_0$ does not involve the fluctuation fields, being given by
\begin{eqnarray}
{\tilde S}_0
&=&
\beta\sum_{\sib p}\xi_{\sib p}
-
{\beta \over 2}\sum_{\sib q}\xi_{B\sib q}
-{\rm Tr}\ln[-{\hat G}(x)^{-1}]
+\beta{{\tilde \Delta}^2 \over U_{\rm eff}}
\nonumber
\\
&=&
\beta\sum_{\sib p}\xi_{\sib p}
-
{\beta \over 2}\sum_{\sib q}\xi_{B\sib q}
-\sum_{{\sib p},\omega_m}{\rm Tr}\ln[-{\hat G}({\bf p},i\omega_m)^{-1}]
+\beta{{\tilde \Delta}^2 \over U_{\rm eff}}.
\label{eq.2.49}
\end{eqnarray}
Here and in the following equations, $G(x)$ is given by Eq. (\ref{eq.2.41}), where ${\tilde \Delta}(x)$ is replaced by the saddle point solution ${\tilde \Delta}=\Delta-g_{\rm r}\phi_{\rm m}$. In the last expression, ${\hat G}({\bf p},i\omega_m)$ is given by Eq. (\ref{eq.2.15}) and the trace is over the particle-hole space. (The trace in the first line involves the integrals over both ${\bf r}$ and $\tau$.) In Eq. (\ref{eq.2.48}), $\delta{\tilde S}$ describes fluctuations around $\Delta$ and $\phi_{\rm m}$,
\begin{eqnarray}
\delta{\tilde S}=
\int_0^\beta d\tau\int d{\bf r}
\Bigl[
{|\delta\Delta(x)|^2 \over U}
-{1 \over 2}\delta{\hat \Phi}^\dagger(x){\hat D}^0(x)^{-1}\delta{\hat \Phi}(x)
\Bigr]
+\sum_{n=2}^\infty{(-1)^n \over n}{\rm Tr}
[({\hat G}(x)\delta{\hat {\tilde \Delta}}(x))^n]
,
\label{eq.2.50}
\end{eqnarray}
where $\delta{\hat {\tilde \Delta}}$ has the same form as Eq. (\ref{eq.2.42}), in which $\Delta(x)$ and $\Phi(x)$ are now replaced by the corresponding fluctuation variables $\delta\Delta(x)$ and $\delta\Phi(x)$, respectively. We note that since we expand $\Delta(x)$ and $\Phi(x)$ around the saddle point solutions ($\Delta$, $\phi_{\rm m}$), the first-order terms in $\delta\Delta(x)$ and $\delta\Phi(x)$ vanish. 
\par
We now employ the {\it Gaussian fluctuation approximation} to Eq. (\ref{eq.2.50}). This approximation corresponds to only retaining quadratic terms with respect to the fluctuation fields, giving
\begin{equation}
\delta{\tilde S}\simeq
\int_0^\beta d\tau\int d{\bf r}
\Bigl[
{|\delta\Delta(x)|^2 \over U}
-{1 \over 2}\delta\Phi^\dagger(x){\hat D}^0(x)^{-1}\delta\Phi(x)
\Bigr]
+{1 \over 2}{\rm Tr}
\Bigl[
{\hat G}(x)\delta{\hat {\tilde \Delta}}(x)
{\hat G}(x)\delta{\hat {\tilde \Delta}}(x)
\Bigr]
.
\label{eq.2.51}
\end{equation}
Eq. (\ref{eq.2.51}) involves $\partial^2{\tilde S}/\partial{\tilde \Delta}^2$ while neglects higher order terms $\partial^n{\tilde S}/\partial{\tilde \Delta}^n$ ($n\ge 3$). This point is crucial when we calculate the equation of the total number of particles from the identity $N=-\partial\Omega/\partial\mu$: The thermodynamic potential $\Omega=-T\ln Z$ has the form
\begin{eqnarray}
\Omega
&=&
T{\tilde S}_0
-T\ln
\int{\cal D}\delta\Phi^\dagger{\cal D}\delta\Phi{\cal D}\delta\Delta^\dagger{\cal D}\delta\Delta
e^{-\delta{\tilde S}}
\nonumber
\\
&\equiv&
T{\tilde S}_0+\Lambda(\Delta,\phi_{\rm m},\mu)
.
\label{eq.2.52}
\end{eqnarray}
Since the last term in Eq. (\ref{eq.2.52}) is $O(\partial^2{\tilde S}/\partial{\tilde \Delta}^2$), the derivative of this term with respect to ${\tilde \Delta}$ gives a higher-order correction under the Gaussian fluctuation approximation. In addition, the ${\tilde \Delta}$-derivative of ${\tilde S}_0$ vanishes due to the saddle point condition given in Eq. (\ref{eq.2.46}). As a result, although the saddle point solution ($\Delta,\phi_{\rm m})$ depends on the chemical potential $\mu$, one does not need to keep the contribution in Eq. (\ref{eq.2.35}) in evaluating $N=-\partial\Omega/\partial\mu$, as we mentioned in Section III. 
\par
In order to carry out the functional integrals in $\Lambda$ in Eq. (\ref{eq.2.52}), we divide the fluctuation fields into the real part and imaginary part as $\delta\Delta(x)=\delta\Delta_R(x)-i\delta\Delta_I(x)$ and $\delta\Phi(x)=\delta\Phi_R(x)-i\delta\Phi_I(x)$. Then we can write $\delta{\hat {\tilde \Delta}}(x)=A(x)\tau_1+B(x)\tau_2$, where $A(x)$ and $B(x)$ are real quantities. When we transform $\delta{\tilde S}$ in Eq. (\ref{eq.2.51}) from the $x(=({\bf r},\tau))$-representation to $q(=({\bf q},\nu_n))$-representation, we find
\begin{eqnarray}
\delta{\tilde S}
&=&
{2 \over \beta U}
{\sum_q}'
{\hat X}_q^\dagger
\Bigl[
1+{U \over 2}{\hat \Pi}^0(q)
\Bigr]
{\hat X}_q
-
{g_{\rm r} \over \beta}
{\sum_q}'
\Bigl[
{\hat X}_q^\dagger{\hat \Pi}^0(q){\hat Y}_q+{\hat Y}_q^\dagger\Pi^0(q){\hat X}_q
\Bigr]
\nonumber
\\
&+&{g_{\rm r}^2 \over \beta}
{\sum_q}'
{\hat Y}_q^\dagger{\hat \Pi}^0{\hat Y}_q
-{2 \over \beta}
{\sum_q}'
{\hat Y}_q^\dagger 
{\hat W} {\hat D}^0(q)^{-1} {\hat W}^{-1}
{\hat Y}_q.
\label{eq.2.53}
\end{eqnarray}
Here ${\hat X}_q^\dagger=(\delta\Delta_R(-q),\delta\Delta_I(-q))$, ${\hat Y}_q^\dagger=(\delta\Phi_R(-q),\delta\Phi_I(-q))$. (Note that $\delta\Delta_{R,I}^*(q)=\delta\Delta_{R,I}(-q)$ and $\delta\Phi_{R,I}^*(q)=\delta\Phi_{R,I}(-q)$.) The prime in the $q$-summation means that one does not take $-q$ when one takes $q$, in order to avoid double counting. ${\hat \Pi}^0(q)$, ${\hat W}$ and ${\hat D}^0(q)$ are, respectively, defined in Eqs. (\ref{eq.2.25}), (\ref{eq.2.30}) and (\ref{eq.2.27}). Executing the functional integral with respect to ${\hat X}_q$, we obtain
\begin{eqnarray}
\delta{\tilde S}
&=&
{1 \over \beta}{\sum_q}'{\rm Tr}\ln
\Bigl[
1+{U \over 2}{\hat \Pi}^0(q)
\Bigr]
-
{g_{\rm r}^2 \over \beta}{\sum_q}'
{\hat Y}_q^\dagger
{\hat \Pi}^0(q)
\Bigl[
1+{U \over 2}{\hat \Pi}^0(q)
\Bigr]^{-1}
{\hat \Pi}^0(q)
{\hat Y}_q
\nonumber
\\
&+&{g_{\rm r}^2 \over \beta}
{\sum_q}'
{\hat Y}_q^\dagger{\hat \Pi}^0{\hat Y}_q
-{2 \over \beta}{\sum_q}'
{\hat Y}_q^\dagger 
{\hat W} D^0(q)^{-1} {\hat W}^{-1}
{\hat Y}_q
\nonumber
\\
&=&
{1 \over 2\beta}\sum_q{\rm Tr}
\Bigl[
1+{U \over 2}{\hat \Pi}^0(q)
\Bigr]
-{2 \over \beta}
{\sum_q}'
{\hat Y}_q^\dagger
\Bigl[
{\hat W}{\hat D}^0(q)^{-1}{\hat W}^{-1}
-{g_{\rm r}^2 \over 2}{\hat \Pi}^0(q)
\Bigl[
1+{U \over 2}{\hat \Pi}^0(q)
\Bigr]^{-1}
\Bigr]
{\hat Y}_q.
\nonumber
\\
\label{eq.2.54}
\end{eqnarray}
In obtaining Eq. (\ref{eq.2.54}), we have used the formula
\begin{equation}
\int{\cal D}{\hat X}_q^\dagger{\cal D}{\hat X}_q 
e^{-\sum_q[{\hat X}_q^\dagger{\hat M}_q{\hat X}_q+B_q^\dagger {\hat X}_q+{\hat X}_q^\dagger C_q]}=
{1 \over {\rm det}[{\hat M}]}e^{\sum_q B_q^\dagger M_q^{-1}C_q}.
\label{eq.2.55}
\end{equation}
Carrying out the functional integral with respect to ${\hat Y}_q$ using this formula, the second term in Eq. (\ref{eq.2.52}) is given by
\begin{eqnarray}
\Lambda(\Delta,\phi_{\rm m},\mu)
&=&
{1 \over 2\beta}\sum_q{\rm Tr}\ln
\Bigl[
1+{U \over 2}{\hat \Pi}^0(q)
\Bigr]
\nonumber
\\
&+&
{1 \over 2\beta}\sum_q{\rm Tr}\ln
\Bigl[
-{\hat W}{\hat D}^0(q)^{-1}{\hat W}^{-1}
+{g_{\rm r}^2 \over 2}
{\hat \Pi}^0(q)
\Bigl[
1+{U \over 2}{\hat \Pi}^0(q)
\Bigr]^{-1}
\Bigr]
\nonumber
\\
&=&
{1 \over 2\beta}\sum_q{\rm Tr}\ln[-{\hat D}^0(q)^{-1}]
+{1 \over 2\beta}\sum_q{\rm Tr}\ln
\Bigl[
1+{1 \over 4}[U-g_{\rm r}^2{\hat D}^0(q)^{-1}]{\hat \Xi}^0(q)
\Bigr].
\label{eq.2.56}
\end{eqnarray}
The matrix correlation function ${\hat \Xi}^0({\bf q},i\nu_n)$ is defined in Eq. (\ref{eq.2.29}).
\par
Substituting Eqs. (\ref{eq.2.49}) and (\ref{eq.2.56}) into Eq. (\ref{eq.2.52}), the thermodynamic potential $\Omega$ is given by
\begin{eqnarray}
\Omega
&=&
{{\tilde \Delta}^2 \over U_{\rm eff}}
+\sum_{\sib p}
\Bigl[\xi_{\sib p}
-{1 \over \beta}\sum_{{\sib p},\omega_m}{\rm Tr}
\ln[-{\hat G}({\bf p},i\omega_m)^{-1}]
\Bigr]
\nonumber
\\
&+&
\Bigl[{1 \over 2\beta}\sum_q
{\rm Tr}\ln[-{\hat D}^0(q)^{-1}]
-{1 \over 2}\sum_{\sib q}\xi_{B\sib q}
\Bigr]
\nonumber
\\
&+&
{1 \over 2\beta}\sum_q{\rm Tr}\ln
\Bigl[
1+{1 \over 4}(U-g_{\rm r}^2{\hat D}^0(q)^{-1}){\hat \Xi}^0(q)
\Bigr].
\label{eq.2.57}
\end{eqnarray}
In turn, the total number of particles $N=-\partial\Omega/\partial\mu$ is given by
\begin{eqnarray}
N
&=&
2\phi_{\rm m}^2
+\sum_{\sib p}
\Bigl\{
1+
{1 \over \beta}\sum_{\omega_m}
{\rm Tr}
\Bigl[
{\hat G}(p){\partial {\hat G}(p)^{-1} \over \partial\mu}
\Bigr]
\Bigr\}
-
\sum_{\sib q}
\Bigl\{
1
+
{1 \over 2\beta}\sum_{\nu_n}
{\partial \over \partial\mu}
{\rm Tr}
\Bigl[
{\hat D}^0(q){\partial {\hat D}^0(q)^{-1} \over \partial\mu}
\Bigr]
\Bigr\}
\nonumber
\\
&-&
{1 \over 2\beta}\sum_{{\sib q},\nu_n}
{\partial \over \partial\mu}
{\rm Tr}\ln
\Bigl[
1+{1 \over 4}(U-g_{\rm r}^2{\hat D}^0(q)^{-1}){\hat \Xi}^0(q)
\Bigr]
\nonumber
\\
&=&
2\phi_{\rm m}^2
+\sum_{\sib p}
\Bigl[1-{\xi_{\sib p} \over E_{\sib p}}\tanh{\beta \over 2}E_{\sib p}
\Bigr]
+2\sum_{\sib q}n_{\rm B}(\xi_{B\sib q})
\nonumber
\\
&-&
{1 \over 2\beta}\sum_{{\sib q},\nu_n}
{\partial \over \partial \mu}
{\rm Tr}\ln
\Bigl\{
1+{1 \over 4}
\Bigl[
U-g_{\rm r}^2{\hat D}^0({\bf q},i\nu_n)
\Bigr]{\hat \Xi}^0({\bf q},i\nu_n)
\Bigr\}.
\label{eq.2.58}
\end{eqnarray}
This result is identical to that given in Eq. (\ref{eq.2.36}). The reproduction of the results in Eqs. (\ref{eq.2.17}) and (\ref{eq.2.36}) [see Eqs. (\ref{eq.2.47}) and (\ref{eq.2.58})] explicitly proves that the strong-coupling theory presented in Section III in terms of summing up diagrams corresponds to keeping Gaussian fluctuations around the (mean-field) order parameters $\Delta$ and $\phi_{\rm m}$. 
\par
\vskip3mm
\section{Goldstone modes in the BCS-BEC crossover Region}
\vskip2mm
\subsection{Correlation function and $b$-boson Green's function in the HF-RPA}
\vskip2mm
The Goldstone mode in fermion superfluidity (Anderson-Bogoliubov) is a collective phase oscillation (phason) of Cooper-pairs, and thus it appears as a pole in the phase correlation function $\Pi_{22}({\bf q},i\nu_n\to\omega+i\delta)$. In the present coupled fermion-boson model, we also expect a Bogoliubov phonon mode associated with the BEC of $b$-molecules characterized by the Bose order parameter $\phi_{\rm m}=\langle b_{{\sib q}=0}\rangle$. This mode appears in the excitation spectrum of the $b$-boson Green's function $D({\bf q},\tau)=-\langle T_\tau\{b_{\sib q}(\tau)b_{\sib q}^\dagger(0)\}\rangle$. However, since the Cooper-pair amplitude $\Delta$ and the $b$-boson order parameter $\phi_{\rm m}$ are coupled with each other via the Feshbach resonance [see Eq. (\ref{eq.2.5})], these two Goldstone modes are strongly hybridized. 
\par
In calculating the Goldstone mode, we have to be careful use a consistent approximation for the self-energy and vertex correction. In this regard, apart from the chemical potential, we recall that the gap equation in Eq. (\ref{eq.2.17}) is obtained from the Hartree-Fock Green's function in Eq. (\ref{eq.2.15}). Thus we should employ the HF-RPA formalism for the correlation functions ${\hat \Pi}=\{\Pi_{ij}\}$ and $b$-boson Green's function\cite{Ranninger2,OhashiT1,OhashiT2}. Using the HF-RPA guarantees the gapless behavior of the Goldstone mode.
\par
In calculating the correlation function ${\hat \Pi}$, it is convenient to first consider ${\hat \Xi}=2{\hat W}^{-1}{\hat \Pi}{\hat W}$ [see Eq. (\ref{eq.2.29})]. Within the HF-RPA, ${\hat \Xi}$ is described by the sum of the diagrams in Fig. 5(a). The summation gives
\begin{eqnarray}
{\hat \Xi}
&=&{\hat \Xi}_U
\Bigl[
1-{1 \over 4}g_{\rm r}^2{\hat D}^0{\hat \Xi}_U
\Bigr]^{-1}
=
{\hat \Xi}^0
\Bigl[
1+
{1 \over 4}
\Bigl[
U-g_{\rm r}^2{\hat D}^0
\Bigr]{\hat \Xi}^0
\Bigr]^{-1}
,
\label{eq.3.1}
\end{eqnarray}
where ${\hat \Xi}_U$ involves the effect of non-resonant interaction $-U$ using the HF-RPA, as shown in Fig. 5(c) and given by Eq. (\ref{eq.2.31b}). The correlation function ${\hat \Pi}$ is then obtained from the inverse relation ${\hat \Pi}={1 \over 2}{\hat W}{\hat \Xi}{\hat W}^{-1}$, namely
\begin{equation}
{\hat \Pi}={\hat \Pi}^0
\Bigl[
1+{1 \over 2}\Bigl[U-g_{\rm r}^2{\hat W}{\hat D}^0{\hat W}^{-1}\Bigr]
{\hat \Pi}^0
\Bigr]^{-1}.
\label{eq.3.2}
\end{equation}
More explicitly, the amplitude and phase correlation functions $\Pi_{11}$ and $\Pi_{22}$ are given by
\begin{eqnarray}
\left\{
\begin{array}{l}
\displaystyle
\Pi_{11}({\bf q},i\nu_n)=
{
{\tilde \Pi}_{11}
\over
\displaystyle
1+{V_1 \over 2}{\tilde \Pi}_{11}
+{V_2^2 \over 4}
{\Pi^0_{11}\Pi^0_{22}-\Pi^0_{12}\Pi^0_{21} \over 1+{V_1 \over 2}\Pi^0_{22}}
-V_2
{\Pi^0_{12} \over 1+{V_1 \over 2}\Pi^0_{22}}
},
\\
\displaystyle
\Pi_{22}({\bf q},i\nu_n)=
{
{\tilde \Pi}_{22}
\over
\displaystyle
1+{V_1 \over 2}{\tilde \Pi}_{22}
+{V_2^2 \over 4}
{\Pi^0_{11}\Pi^0_{22}-\Pi^0_{12}\Pi^0_{21} \over 1+{V_1 \over 2}\Pi^0_{11}}
-V_2
{\Pi^0_{12} \over 1+{V_1 \over 2}\Pi^0_{11}}
},
\end{array}
\right.
\label{eq.3.3}
\end{eqnarray}
where
\begin{eqnarray}
\left\{
\begin{array}{l}
\displaystyle
{\tilde \Pi}_{11}\equiv\Pi^0_{11}
+\Pi^0_{12}
{-{V_1 \over 2} \over 1+{V_1 \over 2}\Pi^0_{22}}
\Pi^0_{21},
\\
\displaystyle
{\tilde \Pi}_{22}\equiv\Pi^0_{22}
+\Pi^0_{21}
{-{V_1 \over 2} \over 1+{V_1 \over 2}\Pi^0_{11}}
\Pi^0_{12},
\end{array}
\right.
\label{eq.3.4}
\end{eqnarray}
\begin{eqnarray}
\left\{
\begin{array}{l}
\displaystyle
V_1({\bf q},i\nu_n)\equiv Re[U_{\rm eff}({\bf q},i\nu_n)]=U+g_{\rm r}^2{\xi_{B\sib q} \over \nu_n^2+\xi_{B\sib q}^{2}},
\\
\displaystyle
V_2({\bf q},i\nu_n)\equiv Im[U_{\rm eff}({\bf q},i\nu_n)]=g_{\rm r}^2{\nu_n \over \nu_n^2+\xi_{B\sib q}^{2}}.
\end{array}
\right.
\label{eq.3.5}
\end{eqnarray}
Here the lowest-order non-interacting correlation functions $\Pi^0_{ij}$ are defined in Eqs. (\ref{eq.2.21a})-(\ref{eq.5.2}). $V_1$ describes amplitude-amplitude and phase-phase interactions, while $V_2$ is an amplitude-phase coupling mediated by $b$-molecules. Since both $V_2$ and $\Pi^0_{12}$ vanish at ${\bf q}=i\nu_n=0$, the denominator of $\Pi_{22}$ in Eq. (\ref{eq.3.3}) reduces to $1+{U_{\rm eff} \over 2}\Pi^0_{22}(0,0)$. This vanishes when the gap equation in Eq. (\ref{eq.2.21d}) is satisfied below $T_{\rm c}$. This proves that the collective phase oscillation is always {\it gapless} at ${\bf q}=0$, a requirement of the Anderson-Bogoliubov (Goldstone) mode.
\par
The $b$-boson Green's function consistent with treating ${\hat \Pi}$ in the HF-RPA is shown in terms of diagrams in Fig. 5(b). The result is just the same as the renormalized $b$-boson Green's function ${\hat D}$ given in Eq. (\ref{eq.2.61}). As discussed at the end of Section III, the excitation spectrum of ${\hat D}$ is {\it gapless} at ${\bf q}=0$ in the superfluid phase below $T_{\rm c}$, which is again consistent with the expected gapless Bogoliubov phonon mode.
\par
\vskip3mm
\subsection{Goldstone mode}
\vskip2mm
As discussed in the preceding subsection, the excitation spectrum of the renormalized $b$-boson is determined by 
\begin{equation}
{\rm det}[{\hat D}({\bf q},i\nu_n\to\omega+i\delta)^{-1}]=0. 
\label{eq.3.6}
\end{equation}
On the other hand, the collective phase oscillation and amplitude oscillation are obtained from the poles of ${\hat \Pi}$ as given by Eq. (\ref{eq.3.2})
\begin{eqnarray}
0
&=&{\rm det}
\Bigl[
1+{1 \over 2}\Bigl[U-g_{\rm r}^2{\hat W}{\hat D}^0{\hat W}^{-1}\Bigr]
{\hat \Pi}^0
\Bigr]_{i\nu_n\to\omega_+=\omega+i\delta}
\nonumber
\\
&=&
{\rm det}[{\hat D}^0({\bf q},\omega_+)]
{\rm det}
\Bigl[
1+{U \over 2}{\hat \Pi}^0({\bf q},\omega_+)
\Bigr]
{\rm det}
\Bigl[
{\hat D}^0({\bf q},\omega_+)^{-1}-{g_{\rm r}^2 \over 4}{\hat \Xi}_U
({\bf q},\omega_+)
\Bigr]
\nonumber
\\
&=&
{
{\rm det}
\Bigl[
1+{U \over 2}{\hat \Pi}^0({\bf q},\omega_+)
\Bigr]
\over
\xi_{B\sib q}^{2}-\omega_+^2
}
{\rm det}[{\hat D}({\bf q},\omega_+)^{-1}].
\label{eq.3.7}
\end{eqnarray}
Comparing Eqs. (\ref{eq.3.6}) with (\ref{eq.3.7}), we see that the correlation functions $\Pi_{ij}$ ($i,j=1,2$) and the $b$-boson Green's function ${\hat D}$ have the identical poles (unless ${\rm det}[1+{U \over 2}{\hat \Pi}^0({\bf q},\omega_+)]$ has zeros). This equivalence is due to hybridizing effects from the coupling between the amplitude fluctuations, phase fluctuations and the $b$-bosons with the amplitude-phase coupling $\Pi^0_{12}$ and the Feshbach resonance coupling $g_{\rm r}$. Thus, in principle, we can consider either one of the correlation functions $\Pi_{ij}$ or the $b$-boson Green's function ${\hat D}$ when we want to study the Goldstone mode. We work with the phase correlation function $\Pi_{22}$ in Eq. (\ref{eq.3.3}), in which case the dispersion relation of the collective mode is given by
\begin{equation}
1+{V_1 \over 2}{\tilde \Pi}_{22}
+{V_2^2 \over 4}
{\Pi^0_{11}\Pi^0_{22}-\Pi^0_{12}\Pi^0_{21} \over 1+{V_1 \over 2}\Pi^0_{11}}
-V_2
{\Pi^0_{12} \over 1+{V_1 \over 2}\Pi^0_{11}}=0.
\label{eq.3.8}
\end{equation}
\par
Landau-damping associated with thermally excited fermion quasi-particles leads to the imaginary part of collective oscillations and, apart from $T=0$, we have to look for a complex solution to Eq. (\ref{eq.3.8}). This requires a complicated analysis. In this paper, for simplicity, we only consider the real part of this equation\cite{OhashiT1,OhashiT2},
\begin{equation}
Re
\Bigl[
1+{V_1 \over 2}{\tilde \Pi}_{22}
+{V_2^2 \over 4}
{\Pi^0_{11}\Pi^0_{22}-\Pi^0_{12}\Pi^0_{21} \over 1+{V_1 \over 2}\Pi^0_{11}}
-V_2
{\Pi^0_{12} \over 1+{V_1 \over 2}\Pi^0_{11}}
\Bigr]=0
.
\label{eq.3.8b}
\end{equation}
From this equation, we can obtain real frequencies as approximate solutions. In order to check the validity of this prescription, we also solve another approximate equation for the mode energy obtained from the renormalized $b$-boson Green's function, 
\begin{equation}
Re[{\hat D}_{11}^0({\bf q},i\nu\to\omega_+)^{-1}]=0.
\label{eq.3.8c}
\end{equation}
As for the damping of the collective mode, we investigate this effect by examining the width of the peak of the collective mode in the spectrum of the correlation functions given by $Im[\Pi_{11}({\bf q},\omega_+)]$ and $Im[\Pi_{22}({\bf q},\omega_+)]$, as well as the $b$-boson excitation spectrum $Im[{\hat D}_{11}({\bf q},i\nu_n\to\omega_+)]$. The structure function
\begin{equation}
S_{jj}({\bf q},\omega)\equiv-{1 \over \pi}[n_{\rm B}(\omega)+1]
Im[{\hat \Pi}_{jj}({\bf q},i\nu\to\omega_+)]~~~~~(j=1,2).
\label{eq.3.9}
\end{equation}
is more convenient than $Im[{\hat \Pi}_{jj}]$ in studying collective behavior, because a diffusive mode spectrum which arises appears as a central peak at $\omega=0$ in $S_{jj}$, which can be easily distinguished from collective modes appearing at a finite frequency. In contrast, this diffusive mode shows up as a peak at a {\it finite} frequency $\omega_d$ in $Im[{\hat \Pi}_{jj}]$, which exhibits a structure of the kind $\omega/(\omega^2+\omega^2_d)$.
\vskip2mm
\subsection{Goldstone mode at zero temperature}
\vskip3mm
At zero temperature, since Landau-damping is absent for modes below the excitation gap $2{\tilde \Delta}$, we can deal directly with Eq. (\ref{eq.3.8}) in determining the Goldstone mode for frequencies $\omega$ below the excitation gap. In the long wavelength or phonon limit, we take $\omega=v_{\phi}q$ and expand Eq. (\ref{eq.3.8}) up to the quadratic order in terms of $\omega$ and $q$. After some calculation, the velocity of the Goldstone mode $v_\phi$ is given by
\begin{eqnarray}
v_{\phi}={1 \over \sqrt{2m}}
{
\displaystyle
\sqrt{
B+{1 \over U_{\rm eff}^2}{g_{\rm r}^2 \over (2\nu-2\mu)^2}
}
\over
\displaystyle
\sqrt{
A
+\eta^2
{{U_{\rm eff} \over 2} \over 1+{U_{\rm eff} \over 2}\Pi^0_{11}(0,0)}
+
{{2 \over U_{\rm eff}}{g_{\rm r}^2 \over (2\nu-2\mu)^2}
\over
1+{U_{\rm eff} \over 2}\Pi^0_{11}(0,0)
}
\Bigl[
{1+{U \over 2}\Pi^0_{11}(0,0) \over [2\nu-2\mu]U_{\rm eff}}
-
\eta
\Bigr]
}
}.
\label{eq.3.10}
\end{eqnarray}
The factors $A$ and $B$ are obtained from the expansion of $\Pi^0_{22}$ as $\Pi^0_{22}({\bf q},i\nu_n)=\Pi^0_{22}(0,0)+A\nu_n^2+Bq^2/2m$, with the explicit expressions
\begin{eqnarray}
\left\{
\begin{array}{l}
\displaystyle
A={1 \over 4}\sum_{\sib p}{1 \over E_{\sib p}^3},
\\
\displaystyle
B=\sum_{\sib p}
\Bigl[
\varepsilon_{\sib p}{{\tilde \Delta}^2 \over 2E_{\sib p}^5}
+{\xi_{\sib p} \over 4E_{\sib p}^3}
\Bigr].
\end{array}
\right.
\label{eq.3.11}
\end{eqnarray}
Finally, the factor $\eta$ is related to the amplitude-phase coupling $\Pi_{12}^0$, namely
\begin{equation}
\eta\equiv
{\Pi^0_{12}({\bf q},i\nu_n) \over \nu_n}
\Bigr|_{T=\nu_n=0}=-{1 \over 2}\sum_{\sib p}{\xi_{\sib p} \over E_{\sib p}^3}.
\label{eq.3.12}
\end{equation}
The second term in the denominator in Eq. (\ref{eq.3.10}) describes the effect of amplitude-phase coupling (second order in $\eta$), while the second term in the numerator and the third term in the denominator involve the effect of the Feshbach resonance coupling (second order in $g_{\rm r}$).
\par
\vskip2mm
\subsubsection{BCS regime: $2\nu\gg2\varepsilon_{\rm F}$}
\vskip3mm
In the BCS limit ($2\nu\gg 2\mu$), the terms involving the factor $1/(2\nu-2\mu)$ can be neglected in Eq. (\ref{eq.3.10}). In addition, since the region near the Fermi surface dominates just as in ordinary weak-coupling BCS theory, we may take
\begin{equation}
\sum_{\sib p} g(\xi_{\sib p})
\to N(\mu(0))\int_{-\infty}^\infty d\xi g(\xi),
\label{eq.3.13}
\end{equation}
where $N(\mu(T=0))$ is the fermion density of states (DOS) at the Fermi surface. In this approximation, $\eta$ coming from the amplitude-phase correlation function $\Pi^0_{12}$ vanishes, and Eq. (\ref{eq.3.10}) reduces to
\begin{equation}
v_{\phi}
=
{1 \over \sqrt{2m}}\sqrt{B \over A}
={1 \over \sqrt{3}}{\bar v}_{\rm F},
\label{eq.3.14}
\end{equation}
where ${\bar v}_{\rm F}\equiv\sqrt{2m\mu(0)}$. In evaluating $B$, we have approximated $\varepsilon_{\sib p}$ appearing in Eq. (\ref{eq.3.11}) by the Fermi energy $\mu(0)$. Eq. (\ref{eq.3.14}) is the well-known velocity of the Anderson-Bogoliubov phonon in weak-coupling BCS superfluidity\cite{Anderson,Sch,Schon}.
\par
The same result as in Eq. (\ref{eq.3.14}) is given by the pole of the phase correlation function $\Pi_{22}$ in the BCS limit. $\Pi_{22}$ in this limit is given by [taking $\Pi^0_{12}=0$, $V_1=U$ and $V_2=0$ in Eq. (\ref{eq.3.3})]
\begin{equation}
\Pi_{22}({\bf q},\omega_+)
=
{\Pi^0_{22}({\bf q},\omega_+) \over 1+{U \over 2}\Pi^0_{22}({\bf q},\omega_+)}
=
{
{2 \over U}\Pi^0_{22}({\bf q},\omega_+)
\over
\Pi^0_{22}({\bf q},\omega_+)-\Pi^0_{22}(0,0)
}
.
\label{eq.3.15}
\end{equation}
Using $\Pi^0_{22}({\bf q},\omega_+)=\Pi^0_{22}(0,0)-A\omega^2+Bq^2/2m$, we easily find that the poles of Eq. (\ref{eq.3.15}) are phonons with the velocity given by Eq. (\ref{eq.3.14}). This shows that the Goldstone mode in the BCS limit is a pure collective phase oscillation of the Cooper-pair order parameter $\Delta$.
\par
The amplitude-phase coupling $\Pi^0_{12}$ in Eq. (\ref{eq.2.14}) vanishes in the BCS limit when we use Eq. (\ref{eq.3.13}) and retain up to $O(q/p_{\rm F})$, where $p_{\rm F}$ is the Fermi momentum. In this case, the amplitude mode is decoupled from Eq. (\ref{eq.3.8}). This mode can then be obtained from the amplitude correlation function working within the HF-RPA,
\begin{equation}
\Pi_{11}({\bf q},\omega_+)
=
{\Pi^0_{11}({\bf q},\omega_+) \over 1+{U \over 2}\Pi^0_{11}({\bf q},\omega_+)}
=
{
{2 \over U}\Pi^0_{11}({\bf q},\omega_+)
\over \Pi^0_{11}({\bf q},\omega_+)-\Pi^0_{22}(0,0)}
.
\label{eq.3.16}
\end{equation}
In particular, at ${\bf q}=0$, Eq. (\ref{eq.3.16}) reduces to
\begin{eqnarray}
\Pi_{11}(0,\omega_+)
&=&
{\Pi^0_{11}(0,\omega_+) \over UN(\mu(0))}
{1 \over {\rm tan}^{-1}{\omega \over \sqrt{4\Delta^2-\omega^2}}}
{\omega \over \sqrt{4\Delta^2-\omega^2}}
~~~~(\omega\le2\Delta)
\nonumber
\\
&\simeq&
{2\Pi^0_{11}(0,2\Delta) \over \pi UN(\mu(0))}
{\Delta \over \sqrt{4\Delta^2-\omega^2}}
~~~~(\omega\simeq2\Delta).
\label{eq.3.17}
\end{eqnarray}
Thus, although $\omega=2\Delta$ is seen to be a branch cut rather than a pole, a strong peak is expected at the excitation gap $\omega=2\Delta$ in the spectrum of the amplitude correlation function $Im[{\hat \Pi}_{11}]$ and hence $S_{11}({\bf q},\omega)$\cite{Schon,Schmid}.
\par
\vskip2mm
\subsubsection{BEC regime: $\nu<0$}
\vskip3mm
In the BEC regime ($\nu<0$), since the chemical potential $\mu$ approaches $\nu$ as $\nu$ decreases, the effective interaction $U_{\rm eff}$ defined in Eq. (\ref{eq.2.8}) is dominated by the Feshbach contribution, $g_{\rm r}^2/(2\nu-2\mu)\gg U$. On the other hand, the correlation functions $\Pi^0_{ij}$ become less dominant. Since $|\mu|\simeq |\nu|\gg{\tilde \Delta}$, the energy gap is also less important in the excitation spectrum of fermion quasi-particles, so that we can approximate $E_{\sib p}=\sqrt{(\varepsilon_{\sib p}-\mu)^2+{\tilde \Delta}^2}\simeq \varepsilon_{\sib p}+|\nu|$. In this limiting case, Eq. (\ref{eq.3.10}) reduces to\begin{eqnarray}
v_\phi
&=&
{g_{\rm r}{\tilde \Delta} \over \sqrt{8m}}
\Bigl[
\sum_{\sib p}{1 \over (\varepsilon_{\sib p}+|\nu|)^3}
\Bigr]^{1/2}
\nonumber
\\
&=&
{g'_{\rm r}{\tilde \Delta} \over \sqrt{8m}}
\Bigl[
{3\pi \over 32|\nu\varepsilon_{\rm F}|^{3/2}}
\Bigr]^{1/2},
\label{eq.3.18}
\end{eqnarray}
where we have rescaled the Feshbach coupling as $g_{\rm r}\sqrt{N}\to g'_{\rm r}$ in the last expression. In evaluating the ${\bf p}$-summation in Eq. ({\ref{eq.3.18}), we have taken into account the correct energy dependence of the DOS ($\propto\sqrt{\varepsilon_{\sib p}}$) in contrast to the approximation in Eq. (\ref{eq.3.13}). Since $\phi_{\rm m}$ is the dominant contribution to ${\tilde \Delta}$ in the BEC regime, Eq. (\ref{eq.3.18}) is proportional to $\sqrt{N_{\rm B}^{\rm c}}$ (where $N_{\rm B}^{\rm c}=\phi_{\rm m}^2$ is the number of condensed $b$-molecules). 
This dependence on $N_{\rm B}^{\rm c}$ is characteristic of the Bogoliubov phonon mode in a Bose-condensed gas. Thus, Eq. (\ref{eq.3.18}) may be regarded as the velocity of the Bogoliubov phonon associated with a condensate of $b$-molecules. We also note that the expression in Eq. (\ref{eq.3.18}) gives as $\sim 1/|\nu|^{3/4}$ and this approaches zero when $\nu\to-\infty$. In this limit, the superfluidity is described by a BEC of a free gas with $N/2$ bosons having the particle-like excitation spectrum $\omega=q^2/2M$, with no linear (or phonon) component.
\par
The Bogoliubov phonon associated with a Bose condensate of paired fermions has also been discussed in strong-coupling superconductivity\cite{Engelbrecht}, as well as for excitons in optically-excited semiconductors\cite{Cote}. Eq. (\ref{eq.3.10}) reproduces the Bogoliubov phonon velocity in strong-coupling superconductivity\cite{Engelbrecht}. In this case, the Feshbach coupling $g_{\rm r}$ and the $b$-boson are absent, while the non-resonant interaction $-U$ is taken to be strong. Then Eq. (\ref{eq.3.10}) reduces to
\begin{eqnarray}
v_{\phi}
&=&
{1 \over \sqrt{2m}}
B^{1/2}
\Bigl[
A+{U \over 2}{\eta^2 \over 1+{U \over 2}\Pi^0_{11}(0,0)}
\Bigr]^{-1/2}
\nonumber
\\
&\to&
{1\over \sqrt{2m}}
\Bigl[
{B(\Delta=0) \over \eta^2(\Delta=0)}
\sum_{\sib p}{\Delta^2 \over (\varepsilon_{\sib p}+|\mu|^2)^3}
\Bigr]^{1/2}
\nonumber
\\
&=&
{\Delta \over \sqrt{8m|\mu|}},
\label{eq.3.19}
\end{eqnarray}
where we have taken the strong-coupling limit ($U\to\infty$), with $\mu$ large and negative. The last expression can be shown to be equivalent to the Bogoliubov phonon velocity in strong-coupling superconductivity, as discussed in Ref.\cite{Engelbrecht}. Using the strong-coupling expressions, $\Delta=(16/3\pi)^{1/2}\varepsilon_{\rm F}/\sqrt{p_{\rm F}a_s}$ and $\mu=-1/2ma_s^2$\cite{Engelbrecht}, where $a_s$ is an $s$-wave scattering length of the pairing interaction between fermions and $p_{\rm F}$ is the Fermi momentum, Eq. (\ref{eq.3.19}) reduces to a more familiar form,
\begin{equation}
v_\phi=
\Bigl[
{n_{\rm B}U_{\rm B} \over M}
\Bigr]^{1/2}.
\label{eq.3.19b}
\end{equation}
Here $n_{\rm B}\equiv N/2V$ is the number density of Cooper-pair bosons (where $V$ is the volume of a system) and $U_{\rm B}\equiv 4\pi a_{\rm B}/M$ ($M=2m$) is an effective interaction between Cooper-pairs with an $s$-wave scattering length $a_{\rm B}\equiv 2a_s$.
\par
\vskip2mm
\section{BCS-BEC crossover in the superfluid phase}
\vskip3mm
In this Section, we present numerical results for the various thermodynamic parameters and correlation functions discussed in earlier Sections, as one passes through the BCS-BEC crossover region. We give results as a function of the temperature in the superfluid phase ($T<T_{\rm c}$) for different values of the $b$-molecule threshold $2\nu$. We take the Fermi energy $\varepsilon_{\rm F}$, Fermi momentum $p_{\rm F}$ and Fermi velocity $v_{\rm F}$ as the units of energy, momentum and velocity, respectively, where $\varepsilon_{\rm F}$, $p_{\rm F}$, and $v_{\rm F}$ are all in the absence of coupling to $b$-bosons. 
As the unit for the number for particles, we take the total number of Fermi atoms $N$. Since the interactions $U$ and $g_{\rm r}$ always appear with $N$ as $UN$ and $g^2_{\rm r}N$, we rescale them as $UN\to U$ and $g_{\rm r}\sqrt{N}\to g_{\rm r}$. As for the energy cutoff which is necessary in the gap equation in Eq. (\ref{eq.2.17}) and the correlation functions $\Pi^0_{ij}$ in Eqs. (\ref{eq.2.21a})-(\ref{eq.5.2}), we employ the Gaussian cutoff $e^{-(\varepsilon_{\sib p}/\omega_c)^2}$ with $\omega_c=2\varepsilon_{\rm F}$, as in our previous work for $T_{\rm c}$\cite{Ohashi1,Ohashi2,Ohashi3}. 
Since our ``strong-coupling" theory is based on perturbative expansions with respect to $-U$ and $g_{\rm r}$, these coupling terms are assumed to be weak perturbations. For this reason, we take $g_{\rm r}=0.6\varepsilon_{\rm F}$ and $U=0.3\varepsilon_{\rm F}$ in all our numerical calculations. In Ref.\cite{Ohashi3}, we have discussed the BCS-BEC crossover at $T_{\rm c}$ for the case of a very broad\cite{Kokkelman,Milstein} Feshbach resonance ($g_{\rm r}\gg\varepsilon_{\rm F}$).
\par
\vskip2mm
\subsection{Temperature dependence of the order parameter and chemical potential}
\vskip3mm
The self-consistent solutions (${\tilde \Delta},\mu$) of the coupled equations (\ref{eq.2.17}) and (\ref{eq.2.36}) are shown in Figs. 6 and 7. In Fig. 6(a), we find that the temperature dependence of the order parameter ${\tilde \Delta}\equiv\Delta-g_{\rm r}\phi_{\rm m}$
agrees well with the BCS theory (`BCS' in the figure). Since ${\tilde \Delta}$ determines the energy gap of the fermion quasi-particle spectrum as $E_{\sib p}=\sqrt{\xi_{\sib p}^2+{\tilde \Delta}^2}$, the superfluid character at $\nu/\varepsilon_{\rm F}=2$ is found to be the BCS-type.
However, Fig. 6(a) also shows a sizable difference between ${\tilde \Delta}$ and the Cooper-pair amplitude $\Delta=\sum_{\sib p}\langle c_{-{\sib p}\downarrow}c_{{\sib p}\uparrow}\rangle$. This is because the BEC order parameter $\phi_{\rm m}=\langle b_{{\sib q}=0}\rangle$ is always induced below $T_{\rm c}$ due to the Feshbach coupling effect in Eq. (\ref{eq.2.5}), which contributes to ${\tilde \Delta}=\Delta-g_{\rm r}\phi_{\rm m}$. As illustration, when we substitute $\nu/\varepsilon_{\rm F}=2$, $g_{\rm r}/\varepsilon_{\rm F}=0.6$, $U/\varepsilon_{\rm F}=0.3$ and $\mu/\varepsilon_{\rm F}\simeq 0.84$ into Eq. (\ref{eq.2.5}), the contribution of the condensed $b$-boson to ${\tilde \Delta}$ is found to be $-g_{\rm r}\phi_{\rm m}=0.517\Delta$ at $T=0$. This means about one third of the order parameter ${\tilde \Delta}$ is associated with $\phi_{\rm m}$ even in the BCS regime at $\nu/\varepsilon_{\rm F}=2$.
\par
As the threshold energy $2\nu$ is lowered, the Cooper-pair component $\Delta$ becomes less dominant while the condensed $b$-boson component $\phi_{\rm m}$ increases (Fig. 6(a) $\to$ (d)). At the same time, the temperature dependence of ${\tilde \Delta}$ deviates from the weak-coupling BCS theory. At $\nu/\varepsilon_{\rm F}=-2$ in Fig. 6(d), ${\tilde \Delta}$ is dominated by $\phi_{\rm m}$ (although the Cooper-pair component also exists due to the Feshbach coupling effect). In this case, $\phi_{\rm m}$ is well described by the BEC order parameter of $N/2$ atoms in a free Bose gas, given by
\begin{equation}
\phi_{\rm m}^{\rm BEC}=\sqrt{N \over 2}
\sqrt{1-\Bigl( {T \over T_{\rm c}} \Bigr)^{3/2}}.
\label{eq.4.2} 
\end{equation}
We denote $\phi_{\rm m}^{\rm BEC}$ in Fig. 6 as `BEC.'
\par
In the intermediate region of the BCS-BEC crossover, ${\tilde \Delta}$ (and also $\Delta$ and $\phi_{\rm m}$) becomes double-valued near $T_{\rm c}$. For example, in Fig. 6(b), the two self-consistent values for ${\tilde \Delta}$ at $T_{\rm c}=0.213T_{\rm F}$ ($\equiv T_{\rm c}^{\rm L}$, where L stands for lower transition temperature) are zero and $0.14\varepsilon_{\rm F}$. In this case, when one decreases the temperature, ${\tilde \Delta}$ jumps abruptly at $T^{\rm L}_{\rm c}$. This is because the phase transition is suppressed by superfluid particle-particle fluctuations, once the phase transition occurs, the opening up of the fermion quasi-particle excitation gap ($2{\tilde \Delta}$) strongly suppresses these fluctuation effects, which accelerates the increase of ${\tilde \Delta}$. On the other hand, when we raise the temperature from below, since the superfluid fluctuations are suppressed by the excitation gap, we can exceed $T_{\rm c}^{\rm L}$ staying in the superfluid phase up to a higher temperature $(\equiv T_{\rm c}^{\rm H})$. In Fig. 6(b), we see that $T_{\rm c}^{\rm H}=0.215T_{\rm F}$, slightly higher than $T_{\rm c}^{\rm L}$. At $T_{\rm c}^{\rm H}$, ${\tilde \Delta}$ vanishes discontinuously. 
\par
Fig. 7 shows that the chemical potential $\mu$ is decreased as the system approaches the BEC regime ($\nu<0$). The temperature dependence of $\mu$ is found to be weak below $T_{\rm c}$ (except just below $T_{\rm c}$ in the crossover regime) compared with the normal phase. As discussed in Section III, the chemical potential is temperature-independent below $T_{\rm c}$ in both the BCS limit ($\mu=\varepsilon_{\rm F}$) and the BEC limit ($\mu=\nu$). Fig. 7 shows that this feature also holds in the intermediate region of the BCS-BEC crossover except just below $T_{\rm c}$, at least for the model parameters we have chosen.
\par
Now that we have calculated self-consistently the values of the composite order parameter ${\tilde \Delta}$ as function of both $T$ and $2\nu$, we can use the results to discuss the spectrum of the BCS-Bogoliubov single-particle Green's function. The excitation spectrum of the BCS-Bogoliubov quasi-particles is given by 
\begin{equation}
\rho_{\rm F}(\omega)\equiv
-{1 \over \pi}\sum_{\sib p}
Im[{\hat G}_{11}({\bf p},i\omega_m\to\omega+i\delta)],
\label{eq.dos1}
\end{equation}
where ${\hat G}_{11}({\bf p},\omega+i\delta)$ is the (11)-component of Eq. (\ref{eq.2.15}). When the chemical potential $\mu$ is positive, the excitation spectrum in Eq. (\ref{eq.dos1}) has a finite gap ${\tilde \Delta}$ as in BCS theory, 
\begin{eqnarray}
\rho_{\rm F}(\omega,\mu\ge 0)
={3N \over 8\varepsilon_{\rm F}^{3/2}}
&\Bigl[&
\Bigl(
{\omega \over \sqrt{\omega^2-{\tilde \Delta}^2}}+1
\Bigr)
\sqrt{\mu+\sqrt{\omega^2-{\tilde \Delta}^2}}
\nonumber
\\
&+&
\Bigl(
{\omega \over \sqrt{\omega^2-{\tilde \Delta}^2}}-1
\Bigr)
\sqrt{\mu-\sqrt{\omega^2-{\tilde \Delta}^2}}
\Theta(\sqrt{{\tilde \Delta}^2+\mu^2}-\omega)
\Bigr]
\Theta(\omega-{\tilde \Delta}),
\label{eq.dos2}
\end{eqnarray}
where $\Theta(x)$ is the step function. On the other hand, when $\mu$ is negative in the BEC regime, one finds,
\begin{eqnarray}
\rho_{\rm F}(\omega,\mu<0)
={3N \over 8\varepsilon_{\rm F}^{3/2}}
\Bigl(
{\omega \over \sqrt{\omega^2-{\tilde \Delta}^2}}+1
\Bigr)
\sqrt{\mu+\sqrt{\omega^2+{\tilde \Delta}^2}}
\Theta(\omega-\sqrt{{\tilde \Delta}^2+\mu^2})
.
\label{eq.dos3}
\end{eqnarray}
In this case, the Fermi quasiparticle excitation gap is found to be $\sqrt{{\tilde \Delta}^2+\mu^2}$, rather than ${\tilde \Delta}$. This reflects the fact that the threshold energy of free fermion excitations is given by $|\mu|$ when $\mu<0$. In the BEC limit ($\mu\simeq\nu\ll-\varepsilon_{\rm F}$), we have $|\mu|\gg{\tilde \Delta}$ and the excitation gap reduces to $|\mu|$.
\par
Fig. 8(a) shows the excitation spectrum of BCS-Bogoliubov quasi-particles in the BCS-BEC crossover. In the BCS regime ($\nu\ge\varepsilon_{\rm F}$), we find a peak at the excitation edge $\omega={\tilde \Delta}$ in the spectrum. This is the well-known coherence peak discussed in the superconductivity literature\cite{Sch}, and the quasi-particle spectrum is found to be the BCS-type in this regime. This coherence is absent when $\nu\le 0$, where the excitation gap gradually changes from ${\tilde \Delta}$ to $|\mu|$ as the threshold energy $2\nu$ is lowered. Since $\mu\simeq\nu$ in the BEC regime, the energy gap ($\sqrt{\mu^2+{\tilde \Delta}^2}$) becomes larger for lower values of $\nu$ in the BEC regime.
\par
Fig. 8(b) shows the momentum distribution function of Fermi atoms at $T=0$\cite{Sch}, which is given by
\begin{equation}
v_{\bf p}^2\equiv\langle c_{{\sib p}\sigma}^\dagger c_{{\sib p}\sigma}\rangle
={1 \over 2}
\Bigl(
1-{\xi_{\sib p} \over E_{\sib p}}
\Bigr).
\label{eq.dis1}
\end{equation}
Since the energy gap ${\tilde \Delta}$ in $E_{\sib p}=\sqrt{\xi_{\sib p}^2+{\tilde \Delta}^2}$ is larger for smaller values of the threshold energy $2\nu$, the steep decrease of $v_{\bf p}^2$ around $\mu(T=0)$ at $\nu/\varepsilon_{\rm F}=2$ gradually disappears as $2\nu$ is lowered. In addition, the magnitude of $v_{\bf p}^2$ decreases as one approaches the BEC regime due to the decrease of the chemical potential $\mu(T=0)$, which reflects the fact that that most Fermi atoms form $b$-bosons in the BEC regime. 
The quantity $v_{\bf p}$ also enters into the Bogoliubov-transformation\cite{Sch} to BCS quasiparticles
\begin{equation}
\gamma^\dagger_{{\sib p}\uparrow}=
u_{\sib p}c^\dagger_{{\sib p}\uparrow}+
v_{\sib p}c_{-{\sib p}\downarrow},
\label{eq.dis2}
\end{equation}
where $\gamma^\dagger_{{\sib p}\uparrow}$ is a creation operator of a BCS-Bogoliubov quasi-particle and $u^2_{\sib p}\equiv 1-v_{\sib p}^2$.
\vskip3mm
\subsection{Velocity of the Goldstone phonon mode}
\vskip2mm
Fig. 9 shows the velocity of the Goldstone mode $v_{\phi}$ at $T=0$ as obtained from Eq. (\ref{eq.3.10}). In the BCS regime ($\nu\gesim\varepsilon_{\rm F}$), the mode velocity $v_\phi$ agrees with the well-known Anderson-Bogoliubov phonon velocity $v_\phi={\bar v}_{\rm F}/\sqrt{3}$ in Eq. (\ref{eq.3.14}). [At $\nu/\varepsilon_{\rm F}=2$, we obtain ${\bar v}_{\rm F}=\sqrt{2m\mu(0)}=0.92v_{\rm F}$ for $\mu(0)\simeq 0.84\varepsilon_{\rm F}$, which gives $v_\phi=0.53v_{\rm F}$.] As the threshold energy $2\nu$ is lowered, $v_\phi$ decreases sharply and approaches the Bogoliubov phonon mode given by Eq. (\ref{eq.3.18}). 
Fig. 9 indicates that $v_\phi$ is strongly dependent on the threshold energy $2\nu$ in a uniform Fermi gas.
\par
Fig. 10 shows the velocity of the Goldstone mode at finite temperatures, obtained from Eq. (\ref{eq.3.8b}). In Fig. 10(a), $v_\phi$ is found to be well described by the Anderson-Bogoliubov mode in the weak-coupling BCS theory (`BCS' in the figure) in the whole temperature region, as expected for the value $\nu=2\varepsilon_{\rm F}$. On the other hand, $v_\phi$ becomes less than the BCS result for the Anderson-Bogoliubov mode as the threshold energy $2\nu$ is decreased. This is shown in Fig. 10(b). Since the order parameter ${\tilde \Delta}$ vanishes discontinuously due to the fluctuation effect discussed in the previous subsection, $v_\phi$ shows a finite jump at $T_{\rm c}$ in Fig. 10(b).
\vskip3mm
\subsection{Dispersion relation of the Goldstone mode}
\vskip2mm
Fig. 11 shows the dispersion of the Goldstone mode at $T=0.5T_{\rm c}$. In the BCS regime (Fig. 11(a)), the gapless dispersion is convex and is confined below the excitation gap at $2{\tilde \Delta}$. This convex dispersion relation gradually changes to a concave one as one goes from Figs. 11(a) to 11(d). In the BEC limit ($\nu<0$), the dispersion is particle-like ($q^2$), characteristic of free $b$-bosons. Indeed the dispersion in Fig. 11(d) is well described by $\omega=q^2/2M$, except for the region of linear dispersion $\omega=v_\phi q$ around ${\bf q}=0$. 
\par
In the BCS regime, Eq. (\ref{eq.3.8c}) also has a high-energy solution at $\omega/\varepsilon_{\rm F}\simeq2.4$, as shown in Fig. 12. This solution is also obtained at $T_{\rm c}$, as shown in Fig. 13(a). Since the energy of this solution is close to the threshold energy of the excitation spectrum of free $b$-bosons as $\xi_{B{\sib q}=0}=2\nu-2\mu=2.32\varepsilon_{\rm F}$ ($\mu\simeq 0.84\varepsilon_{\rm F}$), this high energy solution is interpreted as an excitation of a $b$-boson.
\par
The high-energy mode solution at $T_{\rm c}$ shown in Fig. 13(a) may be useful as a guide to see how the Goldstone mode changes in the BCS-BEC crossover. Fig. 13(a) shows that the energy of this mode gradually decreases as the threshold energy $2\nu$ is lowered. 
At the same time, other solutions appear around $\omega={\bf q}=0$, as shown for $\nu/\varepsilon_{\rm F}=1.29$ in panel (b). At $\nu/\varepsilon_{\rm F}=1.14$ and below, two finite energy and one gapless excitations combine to produce a single phonon. Then the Goldstone mode dispersion below $T_{\rm c}$ starts to change from convex to concave. In Fig. 11(b), although the dispersion below $2{\tilde \Delta}$ is still convex, characteristic of the Anderson-Bogoliubov mode in BCS superfluidity\cite{OhashiT1}, the dispersion above $2{\tilde \Delta}$ obtained from Eq. (\ref{eq.3.8c}) approaches the concave dispersion obtained at $T_{\rm c}$. 
In Figs. 11(c) and (d), the region where linear dispersion occurs ($\omega=v_\phi q$) is very narrow. The overall behavior is the same as the dispersion relation obtained at $T_{\rm c}$, which is particle-like, $q^2/2M$.
\par
Finally we briefly comment on the approximate Eqs. (\ref{eq.3.8b}) and (\ref{eq.3.8c}). Since the excitation gap $2{\tilde \Delta}$ strongly suppresses the fermion quasi-particle excitations far below $T_{\rm c}$, the intraband terms of $\Pi^0_{ij}$ and also the Landau damping below $2{\tilde \Delta}$ are less important at $T=0.5T_{\rm c}$. 
As a result, Eqs. (\ref{eq.3.8b}) and (\ref{eq.3.8c}) both give good approximations to the solution of Eq. (\ref{eq.3.7}) at $T=0.5T_{\rm c}$, and give almost the same results below $2{\tilde \Delta}$, as shown in Figs. 11(a) and (b). 
On the other hand, the interband terms given by the second lines in Eqs. (\ref{eq.2.21a})-(\ref{eq.5.2}) give rise to Landau damping when $\omega\ge 2{\tilde \Delta}$. Thus, above $2{\tilde \Delta}$, Eqs. (\ref{eq.3.8b}) and (\ref{eq.3.8c}) may have different solutions, because they do not include the imaginary part in the same way. Indeed, the dashed line above $2{\tilde \Delta}$ shown in Fig. 11(b), which is a solution of Eq. (\ref{eq.3.8c}), is not obtained from Eq. (\ref{eq.3.8b}). 
In such a case, a more careful analysis is necessary to obtain the correct dispersion relation. 
However, since the fermion quasi-particles are absent in the BEC limit ($\nu<0$), the Landau damping becomes weak in the BEC regime even near $T_{\rm c}$. Thus Eqs. (\ref{eq.3.8b}) and (\ref{eq.3.8c}) again give almost identical results in the whole energy region shown in Figs. 11(c) and (d).
\vskip3mm
\subsection{Spectral weight and damping of the Goldstone mode}
\vskip2mm
In Sections VI.B and VI.C, we considered the Goldstone mode neglecting the Landau damping. In this section, we evaluate the damping from the width of the collective mode in the structure function $S_{jj}({\bf q},\omega_+)$ in Eq. (\ref{eq.3.9}) as well as in the $b$-boson spectral density
\begin{equation} 
\rho_{\rm B}({\bf q},\omega)\equiv -{1 \over \pi}
Im[{\hat D}_{11}({\bf q},\omega_+)].
\label{eq.4.6}
\end{equation}
\par
In Figs. 14-17, we show the $b$-boson excitation spectrum, as well as the phase ($S_{22}$) and amplitude ($S_{11}$) structure functions for $T$, below $T_{\rm c}$. In Fig. 14, we find that the Anderson-Bogoliubov Goldstone mode does not appear as a visible peak in the spectrum at $T/T_{\rm c}=0.99$. (The inset in Fig. 14 shows that the mode energy is $\omega/\varepsilon_{\rm F}=0.01$ for $q=0.02p_{\rm F}$.) This means that the Anderson-Bogoliubov mode is over-damped near $T_{\rm c}$ because of strong Landau damping from the thermally-excited fermion quasi-particles. We note that $S_{22}({\bf q},\omega)$ and $S_{11}({\bf q},\omega)$ both exhibit strong central peaks at $\omega=0$ for $T/T_{\rm c}=0.99$, which indicates the presence of a large number of thermally-excited fermion quasi-particles expected at this temperature. 
\par
At low temperatures, when the Landau damping becomes weaker, the Anderson-Bogoliubov mode appears as a visible peak in $\rho_{\rm B}({\bf q},\omega)$ and $S_{22}({\bf q},\omega)$ as shown in Fig. 14. The peak width becomes narrower at lower temperatures, reflecting the weaker Landau damping by fermions. (At $T=0$, it becomes a sharp $\delta$-function peak.) The peak position ($\omega/\varepsilon_{\rm F}\simeq 0.021$) at $T/T_{\rm c}=0.5$ agrees well with the dispersion relation shown in Fig. 11(a). 
\par
Since the Anderson-Bogoliubov mode is a collective phase oscillation of the Cooper-pair order parameter $\Delta$, the appearance of this mode in $\rho_{\rm B}({\bf q},\omega)$ indicates the presence of the coupling between the $b$-bosons and the phase fluctuations of $\Delta$. On the other hand, no peak structure is observed except for the central peak at $\omega=0$ in the amplitude structure function $S_{11}$ shown in Fig. 14. Only a slight structure appears at $\omega/\varepsilon_{\rm F}\simeq 0.02$ at $T/T_{\rm c}=0.7$, as shown in part (a) of Fig. 18. (The peak at $\omega/\varepsilon_{\rm F}\simeq 0.07$ in Fig. 18 is the amplitude mode at the edge of interband excitations $\omega=2{\tilde \Delta}$.) This is because the amplitude-phase coupling $\Pi^0_{12}$ is very weak in the BCS regime, so that the collective phase oscillation does not strongly couple into the amplitude fluctuations.
\par
As shown in Fig. 12, Eq. (\ref{eq.3.8c}) also has a high energy solution in the BCS regime at $\nu/\varepsilon_{\rm F}=2$. This solution is the strong resonance in $\rho_{\rm B}({\bf q},\omega)$, as shown in part (b) of Fig. 18.
\par
In Fig. 15, we find a broad peak in $S_{22}$ at $T/T_{\rm c}=0.99$. As shown in the inset in Fig. 15, the peak position is different from the one expected from Eqs. (\ref{eq.3.8b}) and (\ref{eq.3.8c}). However, since this resonance shows gapless behavior, it clearly must be the Goldstone mode. (The difference shown in the inset is due to the Landau damping effect.) Indeed, the peak energy at $T/T_{\rm c}=0.5$ agrees well with the dispersion in Fig. 11(b). As expected, Landau damping from fermions becomes weak as one approaches the BEC regime. When $\nu\le 0$, we can observe the sharp peak structure even near $T_{\rm c}$ in Figs. 16 and 17, and the peak position always agrees well with the dispersion relation given in Fig. 1. The dispersion relation of the Goldstone mode approaches the temperature-independent particle-like one in the BEC regime, and hence the temperature dependence of the peak energy in Fig. 17 is weak.
\par 
We also find in Figs. 15-17 that the Goldstone mode appears in the amplitude structure factor $S_{11}({\bf q},\omega)$, which indicates that the amplitude-phase coupling becomes stronger as $2\nu$ is decreased. Although the amplitude fluctuations described by $\Pi_{11}$ are not important in the BCS regime, we cannot neglect these fluctuations in the BCS-BEC crossover regime.
\par
\vskip3mm
\section{Density-density correlation function}
\vskip3mm
An important problem is how to experimentally observe the Goldstone mode discussed in Section V and VI. In this section, we show that the density-density correlation function $\Pi_{33}$ defined in Eq. (\ref{eq.2.21}) exhibits this mode as a pole. This correlation function can be experimentally probed by many techniques, including two-photon Bragg scattering\cite{B1}.
\par
Since the Goldstone mode is associated with the collective phase fluctuations of the Cooper-pairs, a coupling between density fluctuations and phase fluctuations is necessary in order for the Goldstone mode to appear in the spectrum exhibited by $\Pi_{33}({\bf q},\omega_+)$. In this regard, we note that a phase-density coupling exists in fermion superfluidity because of the presence of the Josephson effect. This coupling is described by $\Pi^0_{23}$ (and $\Pi^0_{32}$) defined in Eq. (\ref{eq.5.1}).
In contrast to the amplitude-phase correlation function $\Pi^0_{12}$, which is very weak in the BCS regime, $\Pi^0_{23}$ is finite even if one works with the approximation in Eq. (\ref{eq.3.13}). Besides $\Pi^0_{23}$, density fluctuations also couple with superfluid fluctuations through amplitude-density coupling $\Pi^0_{13}$ (and $\Pi^0_{31}$) defined in Eq. (\ref{eq.5.2}).
This coupling is not important in the BCS regime, because $\Pi^0_{13}$ vanishes when one uses the approximation in Eq. (\ref{eq.3.13}) and neglects term of order $O((q/p_{\rm F})^2)$). However, as in the case of the amplitude-phase correlation function $\Pi_{12}^0$, this coupling effect becomes stronger as one goes into the BCS-BEC crossover regime. Then the density-density correlation function $\Pi_{33}$ is found to couple with superfluid fluctuations (amplitude and phase) via $\Pi_{23}^0$ and $\Pi_{13}^0$, as shown diagrammatically in Fig. 19. 
\par
The density-density correlation function $\Pi_{33}$ given by the HF-RPA is obtained by extending the method discussed in Section V\cite{WongT}. When we introduce a $3\times3$-matrix correlation function ${\hat \Pi}=\{\Pi_{ij}\}~(i,j=1,2,3)$, which involves the density-density correlation function as the $(33)$-component, we obtain an equation similar to Eq. (\ref{eq.3.2}), namely
\begin{equation}
{\hat \Pi}={\hat \Pi}^0
\Bigl[
1+{1 \over 2}{\bar U}{\hat \Pi}^0
\Bigr]^{-1}.
\label{eq.5.3}
\end{equation}
Here the interaction $3\times 3$-matrix ${\bar U}$ is defined by
\begin{eqnarray}
{\bar U}\equiv
\left(
\begin{array}{ccc}
{\bar U}_{11}& {\bar U}_{12}& 0 \\
{\bar U}_{21}& {\bar U}_{22}& 0 \\
0& 0& 0
\end{array}
\right),
\label{eq.5.4}
\end{eqnarray}
where ${\bar U}_{ij}\equiv \{U-g_{\rm r}^2{\hat W}{\hat D}^0{\hat W}^{-1}\}_{ij}$. If we included a density-density interaction term to our coupled fermion-boson model in Eq. (\ref{eq.2.1}), it would be included as the $(33)$-component in Eq. (\ref{eq.5.4}). The density-density correlation function is obtained from the $(33)$-component of Eq. (\ref{eq.5.3}),
\begin{equation}
\Pi_{33}=\Pi_{33}^0+
{
{\bar \Pi}_{33} 
\over 
1
+{V_1 \over 2}[\Pi_{11}^0+\Pi_{22}^0]
-{V_1 \over 2}[\Pi_{12}^0-\Pi_{21}^0]
+[{V_1^2 \over 4}+{V_2^2 \over 4}][\Pi_{11}^0\Pi_{22}^0-\Pi_{12}^0\Pi_{21}^0]
},
\label{eq.5.5}
\end{equation}
where the numerator is
\begin{eqnarray}
{\bar \Pi}_{33}
&\equiv&
[{V_1^2 \over 4}+{V_2^2 \over 4}]
[\Pi_{31}^0\Pi_{12}^0\Pi_{23}^0-\Pi_{31}^0\Pi_{22}^0\Pi_{13}^0
+\Pi_{32}^0\Pi_{21}^0\Pi_{13}^0-\Pi_{32}^0\Pi_{11}^0\Pi_{23}^0
]
\nonumber
\\
&-&{V_1 \over 2}[\Pi_{31}^0\Pi_{13}^0+\Pi_{32}^0\Pi_{23}^0]
-{V_2 \over 2}[\Pi_{31}^0\Pi_{23}^0-\Pi_{32}^0\Pi_{13}^0].
\label{eq.5.6}
\end{eqnarray}
Each term in Eq. (\ref{eq.5.6}) involves correlation functions between density fluctuations and superfluid fluctuations, such as $\Pi_{23}^0$ and $\Pi_{13}^0$.
\par
Fig. 20 shows the dynamic structure function related to the spectral density of density-density correlation function,
\begin{equation}
S_{33}({\bf q},\omega)\equiv
-{1 \over \pi}[n_{\rm B}(\omega)+1]Im[\Pi_{33}({\bf q},i\nu_n\to\omega_+)].
\label{eq.5.7}
\end{equation}
At low temperatures in Figs. 20(a) and (b) (BCS regime), and at all the temperatures in Figs. 20(c) and (d) (BEC regime), we can clearly see the Goldstone mode as a resonant peak in $S_{33}({\bf q},\omega)$. The peak position is the same as that in the spectra shown in Figs. 14-17. Since the density-density correlation function $\Pi_{33}$ can be measured more easily than the phase correlation function $\Pi_{22}$, and may be the most useful way of observing the Goldstone mode.
\vskip3mm
\section{Summary and Discussion}
\vskip2mm
In this paper, we have investigated the BCS-BEC crossover in the superfluid phase of a uniform gas of Fermi atoms with a Feshbach resonance. Going past the simple weak-coupling mean-field theory, we included the strong-coupling effect originating from the pairing interaction associated with the Feshbach resonance. We have extended our previous work\cite{Ohashi1,Ohashi2,Ohashi3} at and above $T_{\rm c}$ to the superfluid region below $T_{\rm c}$. We showed that the superfluid order parameter continuously changes from the Cooper-pair amplitude $\Delta=U\sum_{\sib p}\langle c_{-{\sib p}\downarrow}c_{{\sib p}\uparrow}\rangle$ in the BCS regime to the square root of the number of condensed $b$-molecules $\phi_{\rm m}=\langle b_{{\sib q}=0}\rangle$ associated with the Feshbach resonance in the BEC regime, as one lowers the threshold energy $2\nu$ of the Feshbach resonance. In the intermediate regime in the BCS-BEC crossover, superfluidity is described by the composite order parameter ${\tilde \Delta}\equiv\Delta-g_{\rm r}\phi_{\rm m}$\cite{Timmermans2,Holland,Ohashi1}.
\par
The Goldstone mode is one of the most fundamental phenomena in an ordered system with spontaneous breakdown of a continuous symmetry. In this paper, we investigated how the Anderson-Bogoliubov mode, which is the Goldstone mode in BCS superfluidity, changes to the Bogoliubov phonon mode in the BCS-BEC crossover. The velocity of the Goldstone mode $v_\phi$ strongly depends on the threshold energy $2\nu$, and decreases as one approaches the BEC regime. The Anderson-Bogoliubov phonon mode may be a useful way of monitoring the BCS-BEC crossover.
Since it is difficult to strongly modify the strength of the pairing interaction in metallic superconductors, the tunable pairing interaction associated with the Feshbach resonance in Fermi atomic gases gives one a unique tool to clarify the physics in the BCS-BEC crossover region.
\par
We also investigated the damping of the Goldstone mode. The fermion Landau damping of the Goldstone mode due to coupling to fermions becomes weak far below $T_{\rm c}$, reflecting the fact that the thermal excitation of fermion quasi-particles is negligible. This damping effect is always weak in the BEC regime because the system is then dominated by $b$-molecules, with the suppression of the Fermi quasiparticle spectrum. 
Thus, except in a small region near $T_{\rm c}$ in the BCS regime, the Goldstone mode appears as a strong resonance both in the spectrum of the phase correlation function and in the excitation spectrum of $b$-bosons. As a way to experimentally observe this Goldstone mode, we noted that since the amplitude and phase fluctuations couple with the density fluctuations, one can observe this collective mode in the spectrum of the density-density correlation function $\Pi_{33}({\bf q},\omega_+)$. In cigar-shaped trapped Fermi gases, the direct observation of a density fluctuation pulse might be possible, in analogy with the observation of a Bogoliubov phonon in a superfluid Bose gas with a very weak axial trapping potential\cite{Andrev}.
We also note that two photon Bragg scattering experiments provide a convenient way of studying density fluctuations in a trapped atomic gas\cite{B1,B3}. In such experiments, $Im\Pi_{33}({\bf q},\omega)$ is measured directly, rather than $S_{33}({\bf q},\omega)$\cite{B2}.
\par
It is important to remember that our treatment of the BCS-BEC crossover leaves out several important contributions. In our calculation of various response functions, we always ignored the fermion self-energy arising from the coupling to the non-Bose-condensed bosons\cite{Haussmann2}. In the crossover region, these two-particle states are strongly damped. The numerical results given in Section VI and VII show how Landau damping due to fermions decreases as we approach the BEC regime (small or negative values of $\nu/\varepsilon_{\rm F}$). This is simply because more and more the fermions are forming bound states. However, we expect new forms of damping to arise in the BEC region, namely the Landau and Baliaev damping associated with the interaction between Bogoliubov excitations\cite{PS}.
\par
In this paper, we have not considered the effect of a trap. The atomic density profile in the BCS-BEC crossover in a trap was recently investigated within the LDA by the authors at and above $T_{\rm c}$\cite{Ohashi2,Ohashi3}, using the coupled fermion-boson model, and at $T=0$ by Perali et al.\cite{Stingare}, using the strong-coupling BCS model\cite{Bruun1}. 
A trap potential also leads to various discrete low energy collective modes associated with the confined geometry, and it will be interesting to study how these collective oscillations behave in the BCS-BEC crossover. 
This will be the subject of a future paper, but we conclude with a few brief remarks on this. In the BCS limit of our model, one has the extensive theoretical work\cite{Baranov,Mot,PS} on the Cooper-pair condensate modes in trapped Fermi gases. To be specific, we consider the quadrupole mode $\omega_Q=\sqrt{2}\omega_0$ (where $\omega_0$ is the trap frequency of a spherical trap). This is an analogue of the Goldstone phonon mode in a uniform system. If $\omega_Q$ is greater than the effective threshold $2\Delta_{\rm eff}$ for breaking up Cooper-pairs, this mode is damped and has small spectral weight\cite{Mot}. However, as one approaches the BCS-BEC crossover, the spectral weight of the fermion quasi-particles decreases, shifting to the quadrupole mode $\omega_Q$. In the BEC limit, the mode $\omega_Q=\sqrt{2}\omega_0$ is the well known quadrupole oscillation of a trapped Bose gas\cite{PS,String}. In this paper, we have seen that the Goldstone phonon mode in a uniform gas also persists in the BCS-BEC crossover, but the phonon velocity changes. In contrast, the quadrupole mode frequency does not depend explicitly on the interaction strength, having the same frequency $\omega_Q=\sqrt{2}\omega_0$ in both the BCS and BEC limits. What does change is its spectral weight and damping, which could be used as an experimental signature.
\par
\vskip2mm
\acknowledgements
\par
Y. O. would like to thank Professors S. Takada and M. Ueda for useful discussions, and the Japanese Government for financial support while a visiting professor at the University of Toronto. A. G. acknowledges financial support from the NSERC of Canada.
%

%
\begin{figure}
\caption{(a) $t$-matrix approximation for the particle-particle scattering vertex $\Gamma$ at $T_{\rm c}$. The solid line represents the fermion Green's function. The first line in the figure involves the ladder processes by non-resonant interaction $-U$, while the second line involves the Feshbach resonance described by the $b$-boson Green's function $D_0$. 
(b) Fluctuation contribution to the thermodynamic potential $\Omega$ at $T_{\rm c}$. The first line represents the Cooper-channel particle-particle fluctuations associated with the non-resonant interaction $-U$, and the second line describes the effect of the Feshbach resonance coupling $g_{\rm r}$. (c) The shaded bubble includes multi-ladder scattering processes by $-U$. 
}
\label{Fig1} 
\end{figure}

\begin{figure}
\caption{(a) BCS-BEC crossover at the superfluid phase transition temperature $T_{\rm c}$ in a gas of Fermi atoms with a Feshbach resonance, as a function of the threshold parameter $\nu$. We take $g_{\rm r}/\varepsilon_{\rm F}=0.6$, $U/\varepsilon_{\rm F}=0.6$ and $\omega_c/\varepsilon_{\rm F}=2$. `BCS' labels the result in the absence of fluctuation effects, and `BEC' gives $T_{\rm c}$ of a Bose-condensed gas of $N/2$ molecules of mass $M=2m$. (b) The chemical potential $\mu$ at $T_{\rm c}$, shown as a function of $\nu$. These results are from Ref.12.}
\label{Fig2}
\end{figure}

\begin{figure}
\caption{The off-diagonal static mean-field self-energies in the $2\times 2$-matrix fermion Green's function ${\hat G}({\bf p},i\omega_m)$ given in Eq. (3.11). Diagram (a) gives the contribution from the non-resonant interaction $-U$. Diagrams (b) and (c) include the pairing interaction mediated by a Feshbach molecular boson described by the $b$-boson Green's function $D_0({\bf q},i\nu_n)$. In the diagrams (b) and (c), $\tau_\pm\equiv\tau_1\pm i\tau_2$. Since ${\hat G}({\bf p},i\omega)$ in Eq. (3.11) does not have a $\tau_2$-component, a diagram similar to (a), where $\tau_1$ is replaced by $\tau_2$, is absent.  
}
\label{Fig3}
\end{figure}

\begin{figure}
\caption{(a) Fluctuation contribution ($\delta\Omega$) to the thermodynamic potential below $T_{\rm c}$. The bubble shows the correlation function $\Pi_{ij}^0$. (b) Correlation function of amplitude fluctuations ${\bar \Pi}_{11}$ involving a coupling with phase fluctuations described by the response function $\Pi_{22}^0$. The shaded bubble includes multi-scattering processes by $-U$.
Since the $2\times2$-matrix fermion Green's functions $G$ are given in the Nambu representation, $\Pi_{ij}^0$ is formally described as a particle-hole bubble diagram. (In contrast, $\Pi$ is described by a particle-particle bubble diagram in Fig. 1 because the Nambu two-component representation is not used there.) 
}
\label{Fig4}
\end{figure}

\begin{figure}
\caption{The correlation function ${\hat \Xi}({\bf q},i\nu_n)$ is shown in (a) and $b$-boson Green's function ${\hat D}({\bf q},i\nu_n)$ is shown in (b), both within the HF-RPA. The shaded bubble ${\hat \Xi}_U({\bf q},i\nu_n)$ shown in (c) includes RPA-type diagrams involving the non-resonant attractive interaction $-U$.}
\label{Fig5}
\end{figure}

\begin{figure}
\caption{
Temperature dependence of the superfluid order parameter ${\tilde \Delta}\equiv\Delta-g_{\rm r}\phi_{\rm m}$ in the BCS-BEC crossover region. We also show the separate Cooper-pair component $\Delta$ and the Bose-condensed $b$-molecule component $\phi_{\rm m}$. We take $g_{\rm r}/\varepsilon_{\rm F}=0.6$ and $U/\varepsilon_{\rm F}=0.3$, while ${\tilde \Delta}$ and $\Delta$ are normalized with respect to $\varepsilon_{\rm F}$. The character of the superfluidity changes from the BCS-type to the BEC-type as one goes from (a) to (d). In this figure, `BCS' labels the order parameter given by the weak-coupling BCS theory which omits the particle-particle fluctuations. `BEC' labels the order parameter $\phi_{\rm m}$ of a free Bose gas of $N/2$ atoms, as given by Eq. (6.1).
}
\label{Fig6}
\end{figure}

\begin{figure}
\caption{
The chemical potential $\mu$ as a function of temperature, for $g_{\rm r}/\varepsilon_{\rm F}=0.6$ and $U/\varepsilon_{\rm F}=0.3$. The solid circles show the superfluid phase transition temperature.
}
\label{Fig7}
\end{figure}

\begin{figure}
\caption{
(a) Density of states of BCS-Bogoliubov quasi-particles at $T=0$. For $\nu/\varepsilon_{\rm F}\ge 1$, the threshold energy of the quasiparticle excitations is given by $\omega={\tilde \Delta}$, at which a coherence peak appears. For $\nu/\varepsilon_{\rm F}\le 0$, the coherence peak is absent and the spectrum starts from $\omega=(\mu^2+{\tilde \Delta}^2)^{1/2}$. 
The excitation density of states approaches that for a free Fermi gas of atoms $\rho_{\rm F}(\omega)=(3N/4\varepsilon_{\rm F}^{3/2})(\omega+\mu)^{1/2}$ in the high-energy region.
(b) Momentum distribution function $v_{\bf p}^2\equiv \langle c_{{\sib p}\sigma}^\dagger c_{{\sib p}\sigma}\rangle$ of Fermi atoms with $\sigma$-spin component at $T=0$. The solid circle indicates $\mu(T=0)$. 
}
\label{Fig8}
\end{figure}

\begin{figure}
\caption{
Velocity of the Goldstone phonon $v_\phi$ (normalized to the Fermi velocity) at $T=0$ in the BCS-BEC crossover region. 
}
\label{Fig9}
\end{figure}

\begin{figure}
\caption{
Temperature dependence of the Goldstone phonon velocity $v_\phi$, (a) $\nu/\varepsilon_{\rm F}=2$ (b) $\nu/\varepsilon_{\rm F}=1$. `BCS' labels the Anderson-Bogoliubov phonon velocity given by the weak-coupling BCS theory, where we use $T_{\rm c}$ from Fig. 2 and $\mu$ from Fig. 7. 
}
\label{Fig10}
\end{figure}

\begin{figure}
\caption{
Dispersion relation of the Goldstone mode at $T=0.5T_{\rm c}$, for different values of $\nu$. In this figure, the curves labeled as $[T_{\rm c}]$ shows the dispersion relation at $T_{\rm c}$ obtained from Eq. (5.10). In panels (a) and (b), Eqs.(5.9) and (5.10) give almost the same results below $\omega=2{\tilde \Delta}$. At $\nu/\varepsilon_{\rm F}=2$, Eq. (5.10) has another high energy solution around $\omega/\varepsilon_{\rm F}=2.4$, as shown in Fig. 12. 
}
\label{Fig11}
\end{figure}

\begin{figure}
\caption{
The high energy solution of Eq. (5.10) in the case of $\nu/\varepsilon_{\rm F}=2$, for $T=0.5T_{\rm c}$. The lower energy solution is the Goldstone mode shown in Fig. 11(a).
}
\label{Fig12}
\end{figure}

\begin{figure}
\caption{
(a) Excitation spectrum of the renormalized $b$-boson Green's function at $T_{\rm c}$ obtained from Eq. (5.10). At $\nu/\varepsilon_{\rm F}=2$, there is another solution at $\omega={\bf q}=0$ corresponding to the Goldstone mode. At $\nu/\varepsilon_{\rm F}=1.5$ and 1.29, two finite energy modes and one gapless mode are present, two of them near ${\bf q}=\omega=0$. For illustration, we show these low energy modes for $\nu/\varepsilon_{\rm F}=1.29$ in panel (b), where the two low energy solutions come together and disappear at $q/p_{\rm F}\simeq0.074$. At $\nu/\varepsilon_{\rm F}=0$ and 1.14 in panel (a), only one gapless mode is present. 
} 
\label{Fig13}
\end{figure}

\begin{figure}
\caption{
Spectral weight of the renormalized $b$-boson, $\rho_{\rm B}({\bf q},\omega)=-{1 \over \pi}Im[{\hat D}_{11}({\bf q},\omega)]$, and the structure functions $S_{11}({\bf q},\omega)$ and $S_{22}({\bf q},\omega)$. These results are for $\nu=2\varepsilon_{\rm F}$ (in the BCS region) and $q=0.02p_{\rm F}$. The inset shows the dispersion relation of the Goldstone mode at $T/T_{\rm c}=0.99$: (1) and (2) are obtained from Eqs. (5.9) and (5.10), respectively.}
\label{Fig14}
\end{figure}

\begin{figure}
\caption{
Same plots as in Fig. 14, for $\nu=\varepsilon_{\rm F}$ and $q=0.15p_{\rm F}$.
In the inset, the `peak energy' gives the peak position in $S_{22}({\bf q},\omega)$ at $T/T_{\rm c}=0.99$. 
}
\label{Fig15}
\end{figure}

\begin{figure}
\caption{
Same plots as in Fig. 14, for $\nu=0$ (crossover region) and $q=0.3p_{\rm F}$.
}
\label{Fig16}
\end{figure}

\begin{figure}
\caption{
Same plots as in Fig. 14, for $\nu=-\varepsilon_{\rm F}$ (BEC region) and $q=0.3p_{\rm F}$.
}
\label{Fig17}
\end{figure}

\begin{figure}
\caption{
The amplitude structure function $S_{11}({\bf q},\omega)$ is shown in (a) and the $b$-boson spectrum $\rho_{\rm B}({\bf q},\omega)$ is shown in (b), for $\nu/\varepsilon_{\rm F}=2$ (BCS region) and $T/T_{\rm c}=0.7$. In panel (a), the extremely small structure visible at $\omega/\varepsilon_{\rm F}\simeq 0.02$ is due to the Anderson-Bogoliubov mode. The larger peak at $\omega/\varepsilon_{\rm F}\simeq 0.07$ is the Amplitude mode. However, since the excitation gap is also at $2{\tilde \Delta}=0.068$, this mode coincides with the edge of the interband excitations. In panel (b), the peak on the left is the Anderson-Bogoliubov mode, while the broad peak at $\omega/\varepsilon_{\rm F}$ corresponds to the high energy solution shown in Fig. 12. A very small peak at $\omega/\varepsilon_{\rm F}\simeq 0.068$ in panel (b) is located at the excitation gap $2{\tilde \Delta}$.
}
\label{Fig18}
\end{figure}

\begin{figure}
\caption{
The density-density correlation function $\Pi_{33}$ coupled with superfluid amplitude and phase fluctuations described by $\Pi_{ij}$ ($i,j=1,2$) (shaded bubbles). $\Pi^0_{23}$ and $\Pi^0_{32}$ describe phase-density correlations, while $\Pi^0_{13}$ and $\Pi^0_{31}$ give the amplitude-density correlations.
}
\label{Fig19}
\end{figure}

\begin{figure}
\caption{
Spectrum of the dynamic structure function $S_{33}({\bf q},\omega)$ in the BCS-BEC crossover. The momentum values are the same as in Figs. 14-17. The sharp peak structure is the Goldstone phonon mode.
}
\label{Fig20}
\end{figure}

%
%
%

\begin{references}
\bibitem{note1} The $b$-boson kinetic term ($\sum_{\sib q}\xi_{B\sib q}b_{\sib q}^\dagger b_{\sib q}$) in Eq. (\ref{eq.2.13}) is, strictly speaking, a fluctuation term around $\phi_{\rm m}$. However, since the quadratic contribution to $\Omega$ is easily evaluated, it is included in $\Omega_{\rm MF}$.
\end{references}
\end{document}